\documentclass[aps,prb,superscriptaddress,twocolumn,longbibliography]{revtex4-2}
\usepackage[hidelinks]{hyperref}
\usepackage[T1]{fontenc}
\usepackage{inputenc}
\usepackage{lmodern}
\expandafter\let\csname equation*\endcsname\relax
\expandafter\let\csname endequation*\endcsname\relax
\usepackage{amsmath}
\usepackage{esint}
\usepackage{amssymb}
\usepackage{graphicx}
\usepackage[dvipsnames]{xcolor}
\usepackage{natbib}
\usepackage{bm}
\usepackage{bbold}
\usepackage[mathscr]{euscript}
\usepackage[cal=boondoxo]{mathalfa}
\usepackage{comment}
\usepackage{makecell}
\usepackage{bbding}
\usepackage{pifont}

\def\pM{\mathrel{\raise 2pt \hbox{\tiny(}\!\raise 1pt \hbox{+}\settowidth {\dimen03} {+}\hskip-\dimen03 \raise -2.4pt \hbox {$-$} \!\raise 2pt \hbox{\tiny)}}}

\begin{document}

%\title{Hartree-Fock phase diagram of doped WTe$_2$ monolayers}
\title{Excitonic and magnetic phases in doped WTe$_2$ monolayers:\\ a Hartree-Fock approach}

\author{Guillermo Parra-Mart\'inez}
\affiliation{Fundaci\'{o}n IMDEA Nanociencia, C/ Faraday 9, Campus Cantoblanco, 28049 Madrid, Spain}
\author{Daniel Mu\~noz-Segovia}
\affiliation{Department of Physics, Columbia University, New York, New York 10027, USA}
\author{H\'ector Ochoa}
\email{ho2273@columbia.edu} 
\affiliation{Department of Physics, Columbia University, New York, New York 10027, USA}
\author{Jose Angel Silva-Guill\'en}
\email{joseangel.silva@imdea.org} 
\affiliation{Fundaci\'{o}n IMDEA Nanociencia, C/ Faraday 9, Campus Cantoblanco, 28049 Madrid, Spain}

\begin{abstract}
    Transport and local spectroscopy measurements have revealed that monolayers of tungsten ditelluride ($1T'$-WTe$_2$) display a quantum spin Hall effect and an excitonic gap at neutrality, besides becoming superconducting at low electron concentrations. With the aim of studying the competition among different broken-symmetry phases upon electron doping, we have performed extensive Hartree-Fock calculations as a function of electron density and Coulomb interaction strength. At charge neutrality, we reproduce the emergence of a spin density wave and a spin spiral state surrounding a quantum spin Hall insulator at intermediate interaction strengths. For stronger interactions, the spin spiral is disrupted by a state breaking both inversion and time-reversal symmetries (but not their product) before the system becomes a trivial band insulator. With electron doping the quantum spin Hall insulator evolves into an easy-plane ferromagnet due to a Stoner-like instability of the conduction band. This phase competes energetically with the spin spiral state. We discuss how our results may help to interpret past and future measurements. 
    %Finally, we discuss the experimental implications of our results and their possible impact on the superconducting phase.
\end{abstract}

\maketitle

\section{Introduction}

Transition-metal dichalcogenides (TMDs) display a wide range of physical properties owing to their variety of chemical compositions and polytypes found in nature~\cite{roldan_theory_2017,manzeli_2d_2017,choi_2017_recent,huang_2020_recent,siddique_2021_emerging}. 
In group-VI compounds in particular, the $2H$ polytype has been the subject of a thorough study due to its potential technological applications~\cite{roldan_electronic_2014,manzeli_2d_2017,choi_2017_recent,huang_2020_recent}. 
More recently, the $1T$ polytypes have also attracted some attention, as they can exhibit macroscopic quantum phenomena. 
That is the case of $1T'$-WTe$_2$ monolayers, for which several groups have reported edge conduction compatible with the quantum spin Hall (QSH) effect~\cite{fei_edge_2017,tang_quantum_2017,wu_observation_2018} as well as superconductivity upon low electron doping~\cite{fatemi_electrically_2018,sajadi_gate-induced_2018,song_unconventional_2024,Song_etal2025}.

The $1T$ and $1H$ monolayer structures differ in their coordination: while $1H$ is trigonal, $1T$ is octahedral.
The $1T$ structure is typically not stable and suffers a further reconstruction to the so-called $1T'$ structure, where the transition metal forms a zigzag chain defining a two-fold screw rotation axis. 
Although the $1H$ structure is generally more stable than $1T'$, the latter can be synthesized experimentally. 
Moreover, in the case of WTe$_2$ monolayers the $1T'$ structure is, in fact, the most stable.

Qian \textit{et al}. pointed out that this distortion in the $1T'$ structure causes an inversion between chalcogenide-$p$ and metal-$d$ dominated bands close to the intrinsic Fermi level~\cite{qian_quantum_2014}.  
The resulting band structure corresponds to a Dirac semimetal, whose Dirac points are gapped by the spin-orbit coupling (SOC). 
Whether there is a spectral gap between the resulting electron and hole pockets (leading to a QSH insulator) or the spectrum remains gapless depends heavily on the chemical composition of the material. 
In WTe$_2$ monolayers, electron and hole pockets overlap and the theoretical band structure remains semimetallic in the single-particle picture.
In the work by Qian \textit{et al}., this held true even in the G$_0$W$_0$ band structure calculations~\cite{qian_quantum_2014}. 
Nevertheless, the opening -or not- of a small gap depends heavily on the lattice parameters of the relaxed structure and the specific flavour and parameters chosen for the electronic calculations.
For example, a gap has been reported in Heyd-Scuseria-Ernzerhof  hybrid functional calculations~\cite{zheng_on_2016,tang_quantum_2017,xu_2018_electrically,ok_2019_custodial} and in a so-called self-consistent G$_0$W$_0$ calculation~\cite{wu_quasiparticle_2024}.

\begin{figure}[b!] 
   \includegraphics[width=0.45\textwidth]{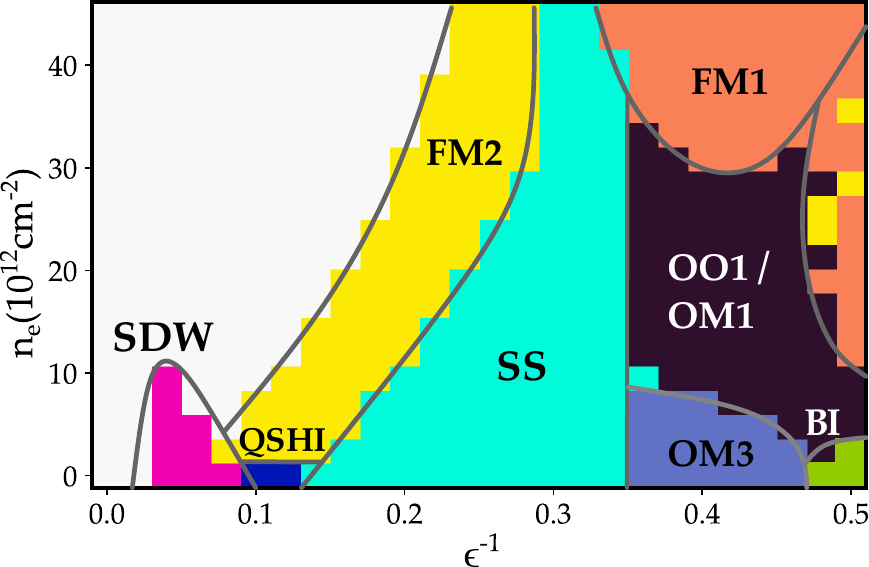}
   \caption{\textbf{Hartree-Fock phase diagram of WTe$_2$ monolayers} at $T=0$. The sequence of solutions at neutrality as a function of interaction strength (horizontal axis) is: a semimetal, an insulator with a spin density wave (SDW), a quantum spin Hall insulator (QSHI), an insulator with a spin spiral (SS), an insulator with orbital and magnetic order (OM3), and a trivial band insulator (BI). Upon doping the QSHI evolves into easy-plane ferromagnet (FM2), which competes in energy with the SS phase. The phase nomenclature along with the order parameters and the (broken) symmetries are summarized in Tab.~\ref{tab:order_parameters}.}
   \label{fig:full}
\end{figure}

Experimentally, although the mechanism for the formation of a gap has been debated for some time (in particular, the role of disorder \cite{song_observation_2018}), recent scanning tunneling microscopy/spectroscopy and transport measurements have confirmed the existence of an excitonic insulating phase~\cite{kohn_excitonic_1967,halperin_possible_1968,halperin_excitonic_1968,volkov_theory_1973} in WTe$_2$ monolayers~\cite{jia_evidence_2022,sun_evidence_2022,que_a-gate-tunable_2024}.
These observations have renewed the interest in exciton physics of topological insulators~\cite{budich_time_2014,pikulin_interplay_2014,xue_time-reversal_2018,blason_exciton_2020,blason_exciton_2020,varsano_monolayer_2020,kwan_theory_2021,amaricci_exciton_2023,wang_breakdown_2023,papaj_spectroscopic_2024} and, more broadly, the role of electron-electron interactions in these materials. 
Coulomb repulsion might drive a change in the character of a topological phase transition~\cite{amaricci_first-order_2015,roy_continuous_2016,amaricci_strong_2016}, as well as stabilize intermediate broken-symmetry states~\cite{xue_time-reversal_2018,blason_exciton_2020,amaricci_exciton_2023}.
In the specific case of WTe$_2$ monolayers, Kwan \textit{et al.} rationalized the absence of the expected charge modulation in the excitonic insulator %associated with the wavevector connecting electron and hole pockets
as the emergence of a spin spiral state at intermediate couplings~\cite{kwan_theory_2021} resulting from an intricate competition fueled by the quantum geometry of the single-electron states. Transport and spectral properties of the spin spirals are in principle compatible with the experiments~\cite{wang_breakdown_2023,papaj_spectroscopic_2024}. Nevertheless, and regardless of the specific form of symmetry breaking, the lack of experimental evidence of a superlattice reconstruction suggests a dominant role of electron-electron interactions over the electron-phonon coupling.

Motivated by these theoretical questions and experimental observations, we have performed extensive Hartree-Fock calculations at zero temperature within a two-band $k\cdot p$ model Hamiltonian of WTe$_2$ including the long-range Coulomb interaction with the aim of elucidating the competition among different broken-symmetry phases and their survival upon electron doping. 
Our findings are summarized in  Fig.~\ref{fig:full}. We obtain a rich phase diagram with several broken-symmetry phases that depend on the two control parameters: the interaction strength given by the (inverse of the) permittivity constant describing the dielectric environment, $\epsilon^{-1}$, and the electron density, $n_e$. Several of these phases are \textit{excitonic}, in the sense that they involve the onset of interband coherences absent in the parent state with no symmetry breaking. The broken-symmetry solutions can be classified in three groups: generalized ferromagnets, density waves, and spin spirals. The last two group of solutions break translational symmetry, but in our model, like in Refs.~\onlinecite{kwan_theory_2021,wang_breakdown_2023}, the spirals do not involve a modulation of the charge density due to an approximate spin quantization axis.

At neutrality, we find a new magnetic state not reported before in the context of WTe$_2$ monolayers. As a function of the interaction strength, we find a sequence of solutions compatible with Ref.~\onlinecite{kwan_theory_2021}: the semimetal described by the $k\cdot p$ model evolves into a spin density wave (SDW) at small couplings, and a spin spiral (SS) over a broad range of intermediate couplings, with a QSH insulating phase in between. However, and contrary to the results in Refs.~\onlinecite{kwan_theory_2021,wang_breakdown_2023}, the SS state does not die off into a trivial band insulator, but instead we find an intermediate magnetic solution involving also orbital order which breaks both inversion and time-reversal symmetries, but not its product. The order parameter describing this solution is the same as in the excitonic phases found between QSH and trivial band insulators in calculations for quantum wells~\cite{xue_time-reversal_2018,blason_exciton_2020,amaricci_exciton_2023}.

At finite doping, our main result is the appearance of an easy-plane ferromagnet due to a Stoner-like instability when electrons are added to the QSH insulator (QSHI); the magnetization lies within the plane perpendicular to the spin quantization axis imposed by SOC. Contrary to the state at neutrality for strong interactions described in the previous paragraph, in this case there is no orbital order involved. This phase competes in energy with the SS state over a broad range of intermediate couplings. This result highlights the possibility that the reported superconductivity arises from (or in proximity to) a state with broken time-reversal symmetry, whose fluctuations might contribute to the pairing glue. 

The rest of the manuscript is structured as follows: In Sec.~\ref{sec:methods}, we describe the model and the numerical details of our calculations. We also introduce the order parameters describing the different phases represented in Fig.~\ref{fig:full}. In Sec.~\ref{sec:results} we analyze the numerical results and construct the phase diagram in a comprehensive way.
Finally, in Sec.~\ref{sec:conclusions} we conclude with a general discussion on how our results can be used to explain features in the experiments. 

\section{Model and calculation scheme}\label{sec:methods}

\subsection{Hamiltonian}

In our calculations we employ a $k\cdot p$ model~\cite{qian_quantum_2014,kwan_theory_2021} describing four electron flavors (two spins in two bands) around the Fermi level. The Hamiltonian $\hat{H}$ consists of two terms. 
One is the single-electron Hamiltonian, $\hat{H}_0=\sum_{\mathbf{k}}\mathcal{H}_{\alpha,\beta}(\mathbf{k})\hat{c}_{\alpha,\mathbf{k}}^{\dagger}\hat{c}_{\beta,\mathbf{k}}$, describing band dispersion with the account of SOC terms. The operator $\hat{c}_{\alpha,\mathbf{k}}^{\dagger}$ ($\hat{c}_{\alpha,\mathbf{k}}$) creates (annihilates) an electron with momentum $\mathbf{k}=(k_x,k_y)$ measured with respect to the zone center and quantum numbers $\alpha=(\tau,s)$ running on the tensor-product space of orbital ($\tau=d,p$) and spin ($s=\uparrow,\downarrow$) degrees of freedom. 
In the $k\cdot p$ approximation the matrix elements $\mathcal{H}_{\alpha,\beta}(\mathbf{k})$ are written as an expansion in powers of $k_i$ constrained by time reversal and the $C_{2h}$ point group symmetries of the $1T'$ structure (space group P2$_1$/m): \begin{align}
\nonumber
  &  \hat{\mathcal{H}}({\mathbf{k}})=\varepsilon_d\left(\mathbf{k}\right)\frac{\hat{1}+\hat{\tau}_z}{2}+\varepsilon_p\left(\mathbf{k}\right)\frac{\hat{1}-\hat{\tau}_z}{2}+\delta\hat{\tau}_z+\hbar v_0k_x\hat{\tau}_y\\
  &  +\hbar v_xk_y\,\hat{\tau}_x\otimes\hat{s}_x+\hbar v_yk_x\,\hat{\tau}_x\otimes\hat{s}_y+\hbar v_zk_y\,\hat{\tau}_x\otimes\hat{s}_z.
  \label{eq:kp_model}
\end{align}
The first line contains the dispersion of a $d_{yz}$-like and a $p_{y}$-like band, where $\varepsilon_d = a\mathbf{k}^2+b\mathbf{k}^4$ and $\varepsilon_p = -\mathbf{k}^2/2m$, respectively. In our convention, the $y$-axis points along the two-fold screw axis of the structure and $z$ is perpendicular to the sample. The parameters $v_0$ and $\delta<0$ describe orbital hybridization and band inversion at the zone center, respectively. The second line contains all the symmetry-allowed SOC terms to linear order in electron momentum. In all these expressions the operators $\hat{\tau}_i$, $\hat{s}_i$ are Pauli matrices acting on orbital and spin quantum numbers, respectively. 

\begin{figure}[t!] 
\includegraphics[width=0.48\textwidth]{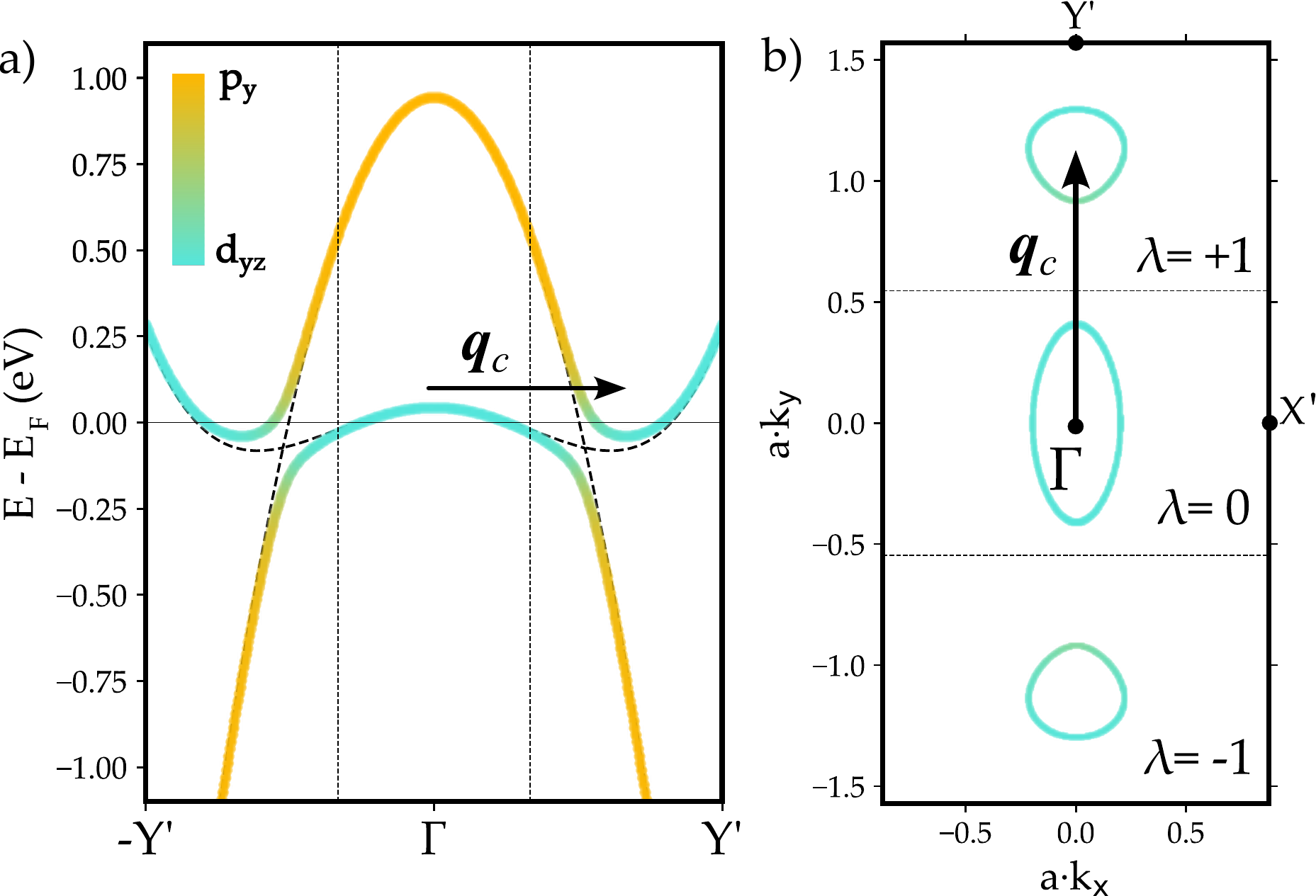}
\caption{\textbf{Non-interacting electron bands and folding scheme}. (a) Non-interacting bands obtained from the $k\cdot p$ model in Eq.~\eqref{eq:kp_model}. The model parameters are: $a=-3$ eV \AA$^2$, $b=18$ eV \AA$^4$, $m=0.03$ eV$^{-1}$ \AA$^{-2}$, $v_x=0.5$ eV \AA, $v_y=3$ eV \AA~and $\delta=-0.45$ eV. 
Black dashed lines represent the bands in the absence of SOC terms. 
Colors show the orbital composition of the bands. 
(b) Fermi contours at neutrality. The lattice constant is $\textrm{a}=3.471$  ($6.221$~\AA\,  along the $y$ axis). Note the slight color change in the cusp of the electron pockets indicating a small contribution of the $p_y$-orbital (same color scale as in the band structure). 
The larger rectangle represents the momentum cutoffs in our calculations. Dashed lines represent zone boundaries of the superlattice.} 
\label{fig:non-interacting}
\end{figure}

Figure~\ref{fig:non-interacting} shows the non-interacting electron band structure deduced from the Hamiltonian in Eq.~\eqref{eq:kp_model}. We use the same model parameters as in Ref.~\onlinecite{jia_evidence_2022,kwan_theory_2021} (see caption of Fig. \ref{fig:non-interacting}). The energies of $d_{yz}$ and $p_y$-dominated bands are inverted at the $\Gamma$ point. In the absence of SOC (black dashed lines in panel a), the $p_y$-like hole-like dispersive band and the much flatter band with dominant $d_{yz}$ character cross at two tilted Dirac points along the $k_y$-axis.
The SOC terms $v_x$ and $v_z$ split these Dirac points, leading to a semimetallic band structure with a hole pocket centered at $\Gamma$ overlaping in energy with two electron pockets located at a distance $\pm\boldsymbol{q}_c$ along the $k_y$-axis. The bands remain two-fold degenerate due to inversion symmetry.

The other term in the Hamiltonian introduces repulsion among electrons by a screened Coulomb interaction,
\begin{align}
\hat{H}_{\textrm{int}}=\frac{1}{2A}\sum_{\mathbf{k}}V\left(\mathbf{k}\right):\hat{N}_{-\mathbf{k}}\hat{N}_{\mathbf{k}}:.
\label{eq:Hint}
\end{align}
Here $\hat{N}_{\mathbf{k}}=\sum_{\alpha,\mathbf{p}}\hat{c}_{\alpha,\mathbf{p}}^{\dagger}\hat{c}_{\alpha,\mathbf{p}+\mathbf{k}}$, $:\,:$ in the equation symbolizes normal ordering of fermion operators, $A$ is the total area of the system, and $V\left(\mathbf{k}\right)$ is the Fourier transform of the Coulomb potential. For simplicity, we assume that the interaction is screened by a double gate in a symmetric device geometry,\begin{align}
V\left(\mathbf{k}\right)=\frac{e^2}{2\,\epsilon\,\epsilon_0 \left|\mathbf{k}\right|}\tanh\frac{\xi \left|\mathbf{k}\right|}{2},
\end{align}
where $e$ is the electron charge, $\epsilon$ is the dielectric constant of the environment, $\epsilon_0$ is the vacuum permittivity constant, and $\xi=72$ nm (which remains fixed in our calculations) represents the separation between gates.

\subsection{Folding scheme}

In the neutral system, the electron-hole band overlap is unstable against the formation of an excitonic gap for some minimum interaction coupling \cite{jia_evidence_2022,varsano_monolayer_2020,kwan_theory_2021}. In order to study if this tendency survives upon electron doping, in our calculations we introduce a folding scheme in which momenta are written as $\mathbf{k}=\mathbf{q}+\lambda\boldsymbol{q}_c$, where $\lambda$ is a new pocket or \textit{valley} quantum number and $\mathbf{q}$ is restricted to the first Brillouin zone of a superlattice with vector $\boldsymbol{q}_c$. The electronic operators are therefore labeled as $\hat{c}_{\alpha,\mathbf{q}+\lambda\boldsymbol{q}_c}\equiv \hat{c}_{\alpha,\lambda,\mathbf{q}}$. The superlattice vector $\boldsymbol{q}_c=(0,0.314)$~\AA$^{-1}$ is defined as the separation along the $y$-axis in momentum space of the centers of the electron and hole pockets in the non-interacting bands and will remain fixed throughout the calculations. In our calculations, momenta $\mathbf{k}$ are restricted to the rectangular area represented in Fig.~\ref{fig:non-interacting}(b), which is half the size of the actual Brillouin zone along $k_x$ and encompasses three superlattice zones in the $k_y$ direction; the valley index runs on $\lambda=0,\pm 1$.

\subsection{Symmetries}

The $k\cdot p$ Hamiltonian in Eqs.~\eqref{eq:kp_model}~and~\eqref{eq:Hint} respects time-reversal $\mathcal{T}$ and the point group $C_{2h}$, generated by inversion $i$ and mirror reflection $\sigma_h$ across the $xz$ plane perpendicular to the screw axis. In terms of the orbital $\hat{\tau}_i$ and spin $\hat{s}_i$ matrices, the representations of $i$, $\sigma_h$ and $\mathcal{T}$ are given by
\begin{subequations}
\begin{align}
    i & :\,\,\hat{\tau}_z,\\
    \sigma_h & :\,\, -i\hat{s}_y,\\
    \mathcal{T} & :\,\, i\hat{s}_y\mathcal{K},
\end{align}
\end{subequations}
where $\mathcal{K}$ is complex conjugation.

The $k\cdot p$ model presents some additional continuous symmetries not contained in the crystalline group. The Hamiltonian $\hat{H}$ is symmetric under continuous translations which, in our folded scheme, translates into a U$_{\lambda}(1)$ symmetry,\begin{align}
    \textrm{U}_{\lambda}(1):\,\, \hat{c}_{\alpha,\lambda,\mathbf{q}}\longrightarrow  e^{-i\lambda\theta}\,\hat{c}_{\alpha,\lambda,\mathbf{q}}.
\end{align}
This should not be confused with a U$(1)$-excitonic symmetry, which refers to separate conservation of electrons and holes and only appears in the \textit{dominant-term} approximation of the model in which inter-band Coulomb scattering processes are neglected. The latter introduce quantum geometry effects through the Coulomb form factors. %\noteDM{link to quantum geometry}. 

It is also customary to take $v_y=0$ in Eq.~\eqref{eq:kp_model}. The consequence is that the model possesses a larger U$_s(1)$ symmetry describing spin rotations along a quantization axis defined by the remaining SOC terms within the $xz$ plane perpendicular to the screw axis. In this approximation the two-fold degeneracy of the non-interacting bands corresponds to up/down spins with respect to this quantization axis. 
The continuous symmetry group of the Hamiltonian is then $G=\textrm{U}_{\lambda}(1)\times\textrm{U}_{s}(1)$.

\subsection{Hartree-Fock equations}

Introducing standard mean-field decouplings in the density (Hartree) and exchange (Fock) interaction channels, $\hat{H}_{\textrm{int}}\rightarrow \hat{H}_{\textrm{H}}+\hat{H}_{\textrm{F}}$, ammounts to the substitution $\hat{H}\rightarrow \hat{H}_{\textrm{HF}}-\langle \hat{H}_{\textrm{H}} \rangle_{\textrm{HF}}-\langle \hat{H}_{\textrm{F}} \rangle_{\textrm{HF}}$, with new single-electron bands described by $\hat{H}_{\textrm{HF}}$, which can be written as the following block-matrix Hamiltonian in the single-electron basis of operators $\hat{c}_{\alpha,\lambda,\mathbf{q}}$:\begin{widetext}\begin{align}
\label{eq:Hartree-Fock_Hamiltonian}
    \hat{\mathcal{H}}_{\textrm{HF}}\left(\mathbf{q}\right)=\left[\begin{array}{ccc}
    \hat{\mathcal{H}}\left(\mathbf{q}\right)+\hat{\Sigma}_{00}\left(\mathbf{q}\right) & \hat{\Sigma}_{0+}\left(\mathbf{q}\right)& \hat{\Sigma}_{0-}\left(\mathbf{q}\right)\\
   \hat{\Sigma}_{0+}^{\dagger}\left(\mathbf{q}\right) & \hat{\mathcal{H}}\left(\mathbf{q}+\mathbf{q}_c\right)+\hat{\Sigma}_{++}\left(\mathbf{q}\right) & \hat{\Sigma}_{+-}\left(\mathbf{q}\right)\\
    \hat{\Sigma}_{0-}^{\dagger}\left(\mathbf{q}\right) & \hat{\Sigma}_{+-}^{\dagger}\left(\mathbf{q}\right) & \hat{\mathcal{H}}\left(\mathbf{q}-\mathbf{q}_c\right)+\hat{\Sigma}_{--}\left(\mathbf{q}\right)
    \end{array}\right].
\end{align}
\end{widetext}
Here the blocks $\hat{\Sigma}_{\lambda_1,\lambda_2}$ represent self-energy matrices in orbital $\otimes$ spin space. Off-diagonal self-energy blocks describe the breaking of U$_{\lambda}$(1) translational symmetry, while diagonal blocks describe exchange renormalizations of the bands that are always important to accurately describe the opening of a gap in the single-electron spectrum~\cite{kwan_theory_2021}. 

Minimization of the free energy for a fixed number of electrons lead to the matrix equations\begin{align}
\label{eq:diagonal}
\hat{\Sigma}_{\lambda\lambda}(\mathbf{q})=-\sum_{\lambda',\mathbf{q}'}V\left(\mathbf{q}-\mathbf{q}'+(\lambda-\lambda')\boldsymbol{q}_c\right)\hat{n}_{\lambda'\lambda'}(\mathbf{q}')
\end{align}
for the block-diagonal terms, and \begin{subequations}
\label{eq:off-diagonal}
\begin{align}
& \hat{\Sigma}_{0\lambda\neq 0}(\mathbf{q}) =V(\boldsymbol{q}_c)\sum_{\mathbf{q}'}\textrm{Tr}\left[\hat{n}_{0\lambda}(\mathbf{q}')+\hat{n}_{-\lambda 0}(\mathbf{q}')\right]\\
& -\sum_{\mathbf{q}'}\left[V\left(\mathbf{q}-\mathbf{q}'\right)\hat{n}_{0\lambda}(\mathbf{q}')
+V\left(\mathbf{q}-\mathbf{q}'+\lambda\boldsymbol{q}_c\right)\hat{n}_{-\lambda0}(\mathbf{q}')\right],
\nonumber\\
& \hat{\Sigma}_{+-}(\mathbf{q})=\sum_{\mathbf{q}'}\left[V(2\boldsymbol{q}_c)\textrm{Tr}\,\hat{n}_{+-}(\mathbf{q}')-V\left(\mathbf{q}-\mathbf{q}'\right)\hat{n}_{+-}(\mathbf{q}')\right]
\end{align}
\end{subequations}
for the off-diagonal block self-energies. In these equations we have density matrices with elements defined by\begin{align}
\label{eq:densities}
\left[\hat{n}_{\lambda_1\lambda_2}(\mathbf{q})\right]_{\alpha\beta}=\frac{1}{A}\left\langle \hat{c}^{\dagger}_{\beta,\lambda_2,\mathbf{q}}\hat{c}_{\alpha,\lambda_1,\mathbf{q}}\right\rangle_{\textrm{HF}},
\end{align}
where the expectation values are computed with respect to the Hartree-Fock Hamiltonian in Eq.~\eqref{eq:Hartree-Fock_Hamiltonian} with the chemical potential set by the equation\begin{align}
\label{eq:chemical_potential}
\sum_{\lambda,\mathbf{q}}\left(\textrm{Tr}\left[\hat{n}_{\lambda\lambda}(\mathbf{q})\right]-\frac{2}{A}\right)=n,
\end{align}
where $n$ is the fixed electron density measured from neutrality. Note that the first terms in the right-hand side of Eqs.~\eqref{eq:off-diagonal} correspond to the Hartree interactions with a non-homogeneous charge distribution within the superlattice. For the block-diagonal self-energies, however, the electrostatic potential associated with the excess charge density $n$ is absorbed in the chemical potential.

\subsection{Order parameters}\label{sec:OP}

Next, we introduce order parameters characterizing the different forms of symmetry breaking in the solutions to the self-consistent Hartree-Fock equations written above. We proceed first to classify the different patterns of symmetry breaking of the continuous symmetry group, $G=\textrm{U}_{\lambda}(1)\times\textrm{U}_{s}(1)$. There are five manifolds in total, divided in three families of solutions: \textit{generalized ferromagnets}, \textit{density waves}, and \textit{spin spirals}. In the first case $\textrm{U}_{\lambda}(1)$ symmetry is preserved, while in the other two cases excitons with momentum $\pm\boldsymbol{q}_c$ condense, although there is no charge modulation associated with the spiral (so long $\textrm{U}_{s}(1)$ is a good symmetry).

\begin{figure}[t!] 
\includegraphics[width=0.35\textwidth]{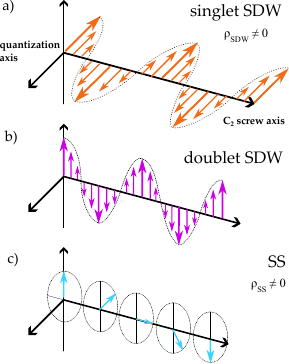}
\caption{\textbf{Spin density waves (SDWs) \textit{vs}. spin spirals (SSs)}. Schematic representation of a singlet SDW (top panel, in orange) polarized along the U$_s$(1) quantization axis within the plane perpendicular to $\boldsymbol{q}_c$, a doublet SDW (middle panel, in purple), and a SS (bottom panel, in blue) rotating within the plane perpendicular to the quantization axis. }
\label{fig:schematics}
\end{figure}

More precisely, the five manifolds of solutions can be defined according to the unbroken subgroup $H\subset G$:\begin{itemize}
\item \textit{Singlet uniform states}, $H=G$, which include the parent state with no symmetry breaking as well as generalized ferromagnets with orbital and/or spin order along the quantization axis imposed by the SOC.
\item \textit{Doublet uniform states}, $H=\textrm{U}_{\lambda}(1)$, which involve spin order within the plane perpendicular to the spin quantization axis. The order-parameter manifold is the circle $G/H=\textrm{U}_{s}(1)$; this phase freedom is associated with a soft magnon.
\item \textit{Singlet waves}, $H=\textrm{U}_{s}(1)$, which in our study can involve a charge density wave (CDW) or a spin density wave (SDW) polarized along the spin quantization axis. These solutions are degenerate in the dominant-term approximation. The order-parameter manifold is the circle $G/H=\textrm{U}_{\lambda}(1)$; this phase freedom is associated with a soft phason.
\item \textit{Doublet waves}, $H=\varnothing$, which in our study consist of a SDW polarized within the plane perpendicular to the spin quantization axis. The order-parameter manifold is the torus $G/H=\textrm{U}_{\lambda}(1)\times\textrm{U}_{s}(1)$; one phase is associated with a soft magnon, the other to a phason.
\item \textit{Spin spirals}, $H=\textrm{U}_{\pm}(1)$, which is a subgroup of $G$ involving admixed translations and spin rotations,\begin{align}
    \textrm{U}_{\pm}(1):\,\, \hat{c}_{(\tau,\tilde{s}),\lambda,\mathbf{q}}\longrightarrow e^{-i(\lambda\pm\frac{\tilde{s}}{2})\theta}\,\hat{c}_{(\tau,\tilde{s}),\lambda,\mathbf{q}},
\end{align}
where $\tilde{s}$ is the polarization along the spin quantization axis. These solutions introduce a superlattice but do not involve a charge modulation. A spin spiral (SS) can be envisioned as a circularly polarized SDW within the plane perpendicular to the quantization axis. The two chiralities (indices $\pm$ above) are energetically degenerate. The order-parameter manifold is $G/H=\textrm{U}_{\mp}(1)$; the phase freedom is associated with a intertwined magnon-phason mode since a translation of the SS is equivalent to a polarization rotation. 
\end{itemize}
Fig.~\ref{fig:schematics} schematically represents the difference between a singlet SDW, a doublet SDW, and a SS.

In order to distinguish these five patterns of symmetry breaking, we introduce the following space-dependent (in general) correlation matrices:\begin{align}
\label{eq:matrix_op}
\rho_{\mu,\nu}=\sum_{\mathbf{q},\lambda_1,\lambda_2}e^{i\left(\lambda_1-\lambda_2\right)\boldsymbol{q}_c\cdot\mathbf{r}}\,\textrm{Tr}\left[\hat{\tau}_{\mu}\otimes\hat{s}_{\nu}\cdot\hat{n}_{\lambda_1,\lambda_2}(\mathbf{q})\right],
\end{align}
where $\mu=0,x,y,z$ run over the identity and the three Pauli matrices in orbital space, and $\nu=0,1,2,3$ run over the identity and the three rotated Pauli matrices in spin space adapted to the quantization axis imposed by SOC.

If we have $\hat{n}_{\lambda_1,\lambda_2}(\mathbf{q})\propto \delta_{\lambda_1,\lambda_2}$ due to U$_{\lambda}(1)$ symmetry, then $\rho_{\mu,\nu}$ is uniform. In singlet uniform states $\rho_{\mu,1}=\rho_{\mu,2}=0$, otherwise it is a doublet solution. Orbital-resolved charge densities $\rho_{0,0}\pm\rho_{z,0}$ are compatible with all the symmetries of the model. The rest of components characterize the spontaneous breaking of some of the discrete symmetries, as indicated in Tab.~\ref{tab:order_parameters}. In the case of singlet states, since the spin quantization axis is not necessarily aligned with a symmetry axis, we find convenient to label all the possibilities with respect to the generators of the magnetic group $C_{2h}(C_i)$ operating within a spin flavor $\tilde{s}$. In the case of doublet states, it is enough to indicate the signature with respect to inversion and time reversal symmetries since a mirror reflection or a screw rotation can be undone by a U$_s(1)$ spin rotation.

\begin{table}
\centering
\begin{tabular}{cccccccccccc}
\hline
\hline
symbol && order parameter && U$_s$(1) && $i$ &&  $\mathcal{T}$  && $\sigma_h\mathcal{T}$  \\
\hline
OO1 && $\rho_{x,0}$ && \ding{51} && \ding{55} && \ding{51} && \ding{51}  \\

OO2 && $\rho_{y,0}$ && \ding{51} && \ding{55} && \ding{55} && \ding{55}   \\

FM1 && $\rho_{0,3}$, $\rho_{z,3}$ && \ding{51} && \ding{51} && \ding{55} && \ding{51} \\

OM1 && $\rho_{x,3}$ && \ding{51} && \ding{55} && \ding{55} && \ding{51}   \\

OM2 && $\rho_{y,3}$ && \ding{51} && \ding{55} && \ding{51} && \ding{55}  \\

FM2  && $(\rho_{0,1},\rho_{0,2})$, $(\rho_{z,1},\rho_{z,2})$ && \ding{55}  && \ding{51} && \ding{55} && -- \\

OM3 && $(\rho_{x,1},\rho_{x,2})$ && \ding{55}  && \ding{55} && \ding{55} && --   \\

OM4 && $(\rho_{y,1},\rho_{y,2})$ && \ding{55}  && \ding{55} && \ding{51} && --  \\
\hline
\hline
\end{tabular}
\caption{\textbf{Order parameters of generalized ferromagnets}. A tick corresponds to a symmetry compatible with a non-zero value of position-independent matrix $\rho_{\mu,\nu}$ defined in Eq.~\eqref{eq:matrix_op}, while a cross implies that the symmetry is necessarily broken in that phase. The phase acronyms stand for \textit{orbital order} (OO), \textit{ferromagnetic} order (FM), and \textit{orbital} and \textit{magnetic} order (OM).}
\label{tab:order_parameters}
\end{table}

When U$_{\lambda}$(1) is broken by valley-off diagonal terms, the correlation matrices are modulated with wavevectors $\boldsymbol{q}_c$, $2\boldsymbol{q}_c$. To characterize the solution we can just focus on the Fourier component $\boldsymbol{q}_c$ of the charge and spin densities, respectively, \begin{subequations}
    \begin{align}
    & \rho_{\boldsymbol{q}_c}=\sum_{\boldsymbol{q}}\textrm{Tr}\left[\hat{n}_{+,0}(\mathbf{q})+\hat{n}_{0,-}(\mathbf{q})\right],\\
    & \rho_{\boldsymbol{q}_c}^{(i)}=\sum_{\boldsymbol{q}}\textrm{Tr}\left[\hat{\tau}_0\otimes\hat{s}_i\left(\hat{n}_{+,0}(\mathbf{q})+\hat{n}_{0,-}(\mathbf{q})\right)\right].
\end{align}
\end{subequations} Since the superlattice is incommensurate with the microscopic lattice, we disregard orbital orders and breaking of point group symmetries in our classification. The order parameter of the CDW and singlet SDW solutions are $\rho_{\textrm{CDW}}\equiv|\rho_{\boldsymbol{q}_c}|$ and $\rho_{\textrm{SDW}_1}\equiv|\rho_{\boldsymbol{q}_c}^{(3)}|$, respectively. To describe doublet SDW and SS solutions, we follow Ref.~\onlinecite{kwan_theory_2021} and define from $\boldsymbol{\rho}_{\boldsymbol{q}_c}=(\rho_{\boldsymbol{q}_c}^{(1)},\rho_{\boldsymbol{q}_c}^{(2)},0)$:\begin{subequations}
    \begin{align}
&\rho_{\textrm{SDW}_2}\equiv\sqrt{2|\boldsymbol{\rho}_{\boldsymbol{q}_c}\cdot\boldsymbol{\rho}_{\boldsymbol{q}_c}|}, \\
& \rho_{\textrm{SS}}\equiv \sqrt{2\boldsymbol{\rho}_{\boldsymbol{q}_c}\cdot\boldsymbol{\rho}_{\boldsymbol{q}_c}^*}-\sqrt{2|\boldsymbol{\rho}_{\boldsymbol{q}_c}\cdot\boldsymbol{\rho}_{\boldsymbol{q}_c}|}.
\end{align}
\end{subequations}

\subsection{Numerical procedure}

\begin{table}
\centering
\begin{tabular}{ccccccc}
\hline
\hline
Non-uniform Ansatz && $\hat{\Sigma}_{0,1}$ && $\hat{\Sigma}_{0,-1}$  \\
\hline

CDW && $\hat{\tau}_r\otimes\hat{s}_0$ && $\hat{\tau}_r\otimes\hat{s}_0$   \\
Singlet - SDW && $\hat{\tau}_r\otimes\hat{s}_3$ && $\hat{\tau}_r\otimes\hat{s}_3$   \\
Doublet - SDW && \makecell{$\hat{\tau}_r\otimes[\textrm{cos}(\theta_r)\hat{s}_1-$\\$\textrm{sin}(\theta_r)\hat{s}_2]$} &&  \makecell{$\hat{\tau}_r\otimes[\textrm{cos}(\theta_r)\hat{s}_1-$\\$\textrm{sin}(\theta_r)\hat{s}_2]$} \\
SS && $\hat{\tau}_r\otimes(\hat{s}_1+i\hat{s}_2)$ && $\hat{\tau}_r\otimes(\hat{s}_1-i\hat{s}_2)$  \\
\hline
\hline
\end{tabular}
\caption{\textbf{Self-energy ansatzs for non-uniform states}. Kronecker products act on $\textrm{orbital } \otimes \textrm{ spin}$ subspaces. $\hat{\tau}_r$ refers to a $2\times2$ random matrix in orbital space and $\theta_r$ to a uniform random angle between $[0,2\pi)$.}
\label{tab:ansatz}
\end{table}

Equations~\eqref{eq:diagonal}~and~\eqref{eq:off-diagonal} together with Eqs.~\eqref{eq:densities}~and~\eqref{eq:chemical_potential} define a complete set of non-linear equations that we solved numerically in an iterative procedure. All calculations were performed on a $27\times255$ Monkhorst-Pack k-point grid within the rectangle in Fig.~\ref{fig:non-interacting}~b (see model parameters in the caption of the same figure). The phase diagram in Fig.~\ref{fig:full} and the analysis presented in the following section are the result of self-consistent calculations for 26 
equispaced values of $\epsilon^{-1}$ up to $\epsilon^{-1}_{\text{max}}=0.5$, and 20 density values up to $n_{\text{max}}=4.547\cdot10^{13}\:\text{cm}^{-2}$. All calculations were performed at zero temperature.

To explicitly look for broken-symmetry solutions in our iterative procedure, we initially break the symmetry by introducing a small momentum-independent self-energy ansatz with suitable structure in orbital, spin, and valley indices. The iteration continues until 
the total energy converges to within $%\Delta E=
0.1\:\mu eV$, which we have checked leads to well-converged solutions. 
Nevertheless, we have checked that our results do not change when we converge the order parameter instead of the energies. 
For uniform solutions, the ansatz was diagonal in valley, and calculations were performed in the extended zone without folding. We tried initial self-energy ansatzs with the same orbital and spin  matrix structure as the order parameters introduced in Tab.~\ref{tab:order_parameters}. 
For non-uniform solutions, we considered valley off-diagonal ansatzs, $\hat{\Sigma}_{0,1}$ and $\hat{\Sigma}_{0,-1}$ (and their hermitian conjugates), and calculations were performed within a folded scheme.
The spin matrix structures of a CDW, SDW, or SS initial ansatz are given in Table \ref{tab:ansatz}.
The entries in orbital space were randomized using the Distributions package in Julia. 

\section{Results}\label{sec:results}

The phase diagram in Fig.~\ref{fig:full} results from the intricate competition between uniform generalized ferromagnets and non-uniform density waves and spirals. 
We first discuss the former group of solutions imposing off-valley self-energy blocks to zero shown in Fig.~\ref{fig:parent}. 
We later compare their energies with the solutions breaking translational symmetry, previously obtained by Kwan \textit{et al}. \cite{kwan_theory_2021}. 
Then we focus on the sequence of ground states at charge neutrality as a function of interaction strength, and we finally construct the full phase diagram in Fig.~\ref{fig:full}, paying special attention to the new ferromagnetic phase that arises from the QSHI upon electron doping.

\subsection{Generalized ferromagnets}
\label{subsec:q=0}

\begin{figure}[b!] 
\includegraphics[width=0.45\textwidth]{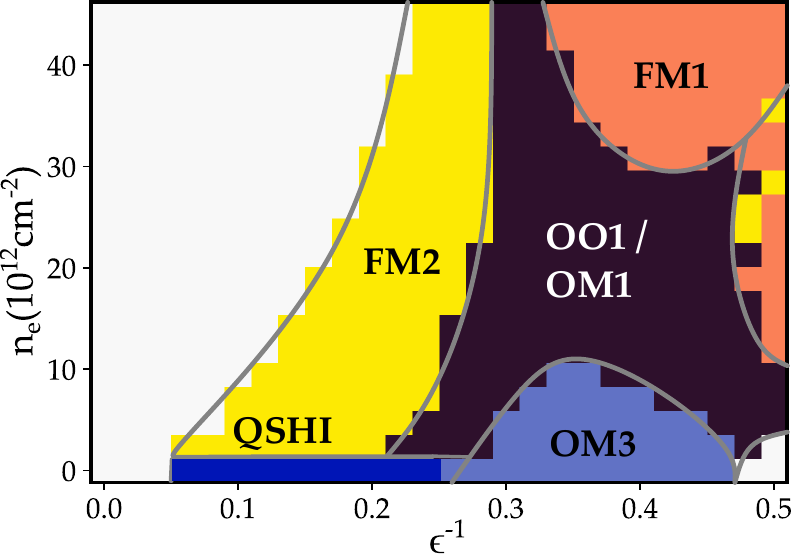}
\caption{\textbf{Lowest-energy solutions preserving translational symmetry}. The FM2 and OM3 phases compete directly with the QSH and SS insulators.}
\label{fig:parent}
\end{figure}

\begin{figure}[t!]
    \centering
\includegraphics[width=1\columnwidth]{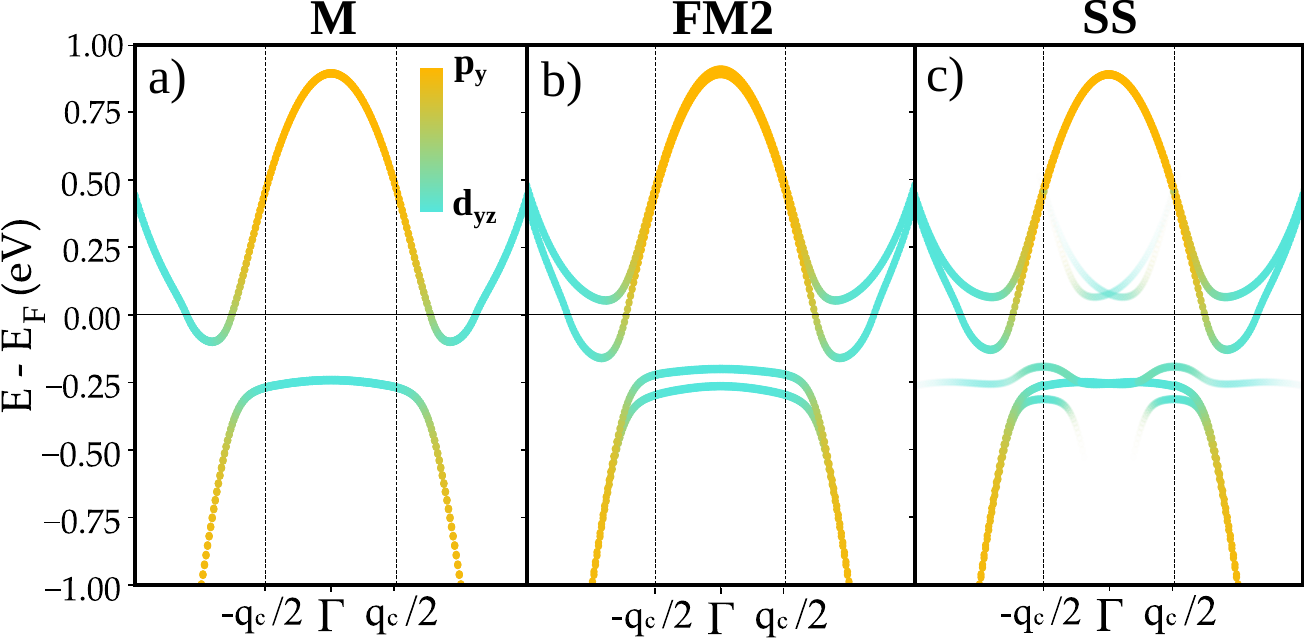}
    \caption{\textbf{Orbital-resolved electron bands} of (a) normal state, (b) FM2, and (c) SS solutions for $\epsilon^{-1}=0.2$ and $n_e=16.75\cdot10^{12}$cm$^{-2}$. Color indicates the orbital composition. In panel (c) the color opacity reflects the spectral weight of the single-electron states of the SS solution after band unfolding.}
    \label{fig:bands}
\end{figure}

Figure~\ref{fig:parent} displays the lowest-energy solutions preserving translational symmetry, i.e., with the valley off-diagonal elements in Eq.~\eqref{eq:Hartree-Fock_Hamiltonian} turned off. Besides the parent state involving no symmetry breaking (non-colored pixels), we find five solutions with different forms of orbital and spin order.

Let us first analyze the evolution of the phase diagram at charge neutrality. For small values of $\epsilon^{-1}$, the system is a compensated semimetal
(see Fig.~\ref{fig:non-interacting}).
By slightly increasing the interaction strength above $\epsilon^{-1}=0.06$, the system develops a
gap by (an almost rigid) relative shift of the electron and hole bands without symmetry breaking. Due to the Berry curvature inherited from the Dirac points gapped by SOC, this leads to the QSHI phase (dark blue in Fig.~\ref{fig:parent}). 
For very large interaction strengths ($\epsilon^{-1}\gtrsim0.47$), the ground state becomes a trivial insulator with the bands inverted at the $\Gamma$ point with respect to the QSHI. However, the direct topological transition between the QSHI and trivial insulators is avoided by an intermediate broken-symmetry phase arising between $\epsilon^{-1}\simeq0.25$ and $\epsilon^{-1}\simeq0.47$ (the blue OM3 phase in Fig.~\ref{fig:parent}). This OM3 phase is an orbital and spin ordered phase with broken U$_s$(1), inversion $i$ and time-reversal symmetries $\mathcal{T}$ (but it preserves the product $i\mathcal{T}$). %), notice the band degeneracy in Fig.~\ref{fig:bands}~a)).
Such an intermediate broken-symmetry phase, which avoids the gap closing in the topological transition, generically appears between two topologically distinct states. Indeed, the OM3 order parameter is analogous to that found in the interacting Bernevig-Hughes-Zhang model for quantum wells~\cite{xue_time-reversal_2018,blason_exciton_2020,amaricci_exciton_2023}.

As soon as we electron dope the QSHI, a ferromagnetic phase develops (yellow FM2 phase in Fig.~\ref{fig:parent}). Remarkably, this is the lowest-energy uniform solution in a broad range of dopings and intermediate interaction strengths. The FM2 phase is an intraorbital ferromagnet whose magnetization lies within the plane perpendicular to the SOC quantization axis, breaking the U$_s(1)$ symmetry. The FM2 can be understood as a Stoner instability of the conduction band. Indeed, as shown in Fig.~\ref{fig:bands}, the strongest effect of the FM2 on the band structure is to spin split the conduction band around the electron pockets close to $\pm q_c$. The FM2 saves band energy by placing its chemical potential in the middle of the spin-split pockets, which competes against the increased exchange energy due to the spin splitting. 
Above a critical doping that increases with the interaction strength, the exchange energy cost is higher than the band energy saved, inducing a first order transition to the parent state, which is a doped QSHI without broken symmetries. 

For interactions stronger than $\epsilon^{-1}\simeq0.25$, the OM3 phase, which is the ground state at charge neutrality, survives until a relatively small electron doping. Further electron doping the system drives a first order transition to the OO1 or OM1 states (black in Fig.~\ref{fig:parent}). These are orbital-ordered spin-singlet states, which are degenerate at the Hartree-Fock level since they are related by inversion acting only on one spin flavor. The transition from the OM3 to the OO1/OM1 state can also be understood as a Stoner instability due to the splitting of the doped conduction band in the OO1/OM1. 
We finally mention that, at higher dopings and interaction strengths, the lowest-energy solution is the spin-singlet ferromagnet FM1 (orange in Fig.~\ref{fig:parent}), whose magnetization points along the SOC spin quantization axis. 

\subsection{Phase diagram at neutrality}

\begin{figure}[b!] 
\includegraphics[width=0.45\textwidth]{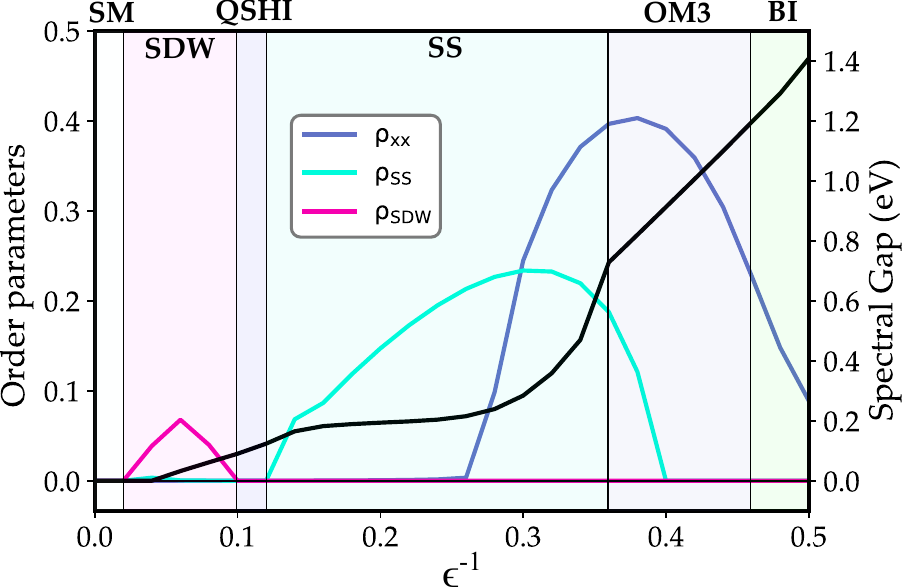}
\caption{\textbf{Order parameters and gap at charge neutrality as a function of interaction strength}. On the left axis we represent the order parameters of the lowest-energy solution breaking translational symmetry: a doublet SDW (in pink) or a SS (cyan), together with the order parameter of the OM3 solution (dark blue). The colored shadowed areas correspond to the ground state indicated by the acronym, from left to right: semimetal (white), SDW (pink), QSHI (blue), SS (light blue), OM3 (dark blue), and trivial band insulator (BI, green). On the right axis we show the spectral gap (black curve). The gap opens within the SDW phase, leading to a QSHI between the SDW and SS phases. At stronger interactions, the first-order transition between SS and OM3 phases is marked by a kink in the gap. The last transition to a trivial BI appears to be weakly first order.} 
\label{fig:symbreak}
\end{figure}

We now study the interplay of generalized ferromagnets with non-uniform states (allowing for valley off-diagonal blocks in Eq.~\ref{eq:Hartree-Fock_Hamiltonian}) at charge neutrality. Our calculations reveal two competing manifolds of solutions breaking translational symmetry: doublet SDWs and SSs. CDWs and singlet SDWs
are either not stable or not energetically competitive. 

Figure~\ref{fig:symbreak} shows the evolution of the order parameters and single-electron band gap at neutrality as a function of interaction strength. 
At weak interactions, a semimetallic SDW develops. As the order parameter grows the system eventually develops a gap. The SDW phase dies around $\epsilon^{-1}\simeq0.1$, leading to a small range of interactions with a QSHI ground state without broken symmetries. At $\epsilon^{-1}\simeq0.12$, the system transitions into the SS state, which is stable over a wide range of interaction strengths. The SS, which has a stronger  $d$-orbital component, is stabilized by a reconstruction of the spectrum that pushes the valence band downward in energy. 
However, the SS solution is no longer energetically favorable at strong interactions owing to the tendency of the system to undo the band inversion of the original semimetallic model at the zone center. This sequence of ground states as a function of $\epsilon^{-1}$ agrees with that reported in Ref.~\onlinecite{kwan_theory_2021}.

However, in contrast to Ref.~\onlinecite{kwan_theory_2021}, we find that there is no direct transition from the SS to a trivial band insulator, but rather there is an intermediate phase that appears above $\epsilon^{-1}\simeq0.35$ corresponding to the OM3 solution discussed before. The transition to the trivial band insulator occurs at a higher interaction strength, $\epsilon^{-1}\simeq0.45$. Both the SS to OM3 and OM3 to trivial band insulator transitions are first-order, characterized by a kink in the spectral gap and a discontinuity of the order parameters.

\begin{figure}[t!] 
\includegraphics[width=0.48\textwidth]{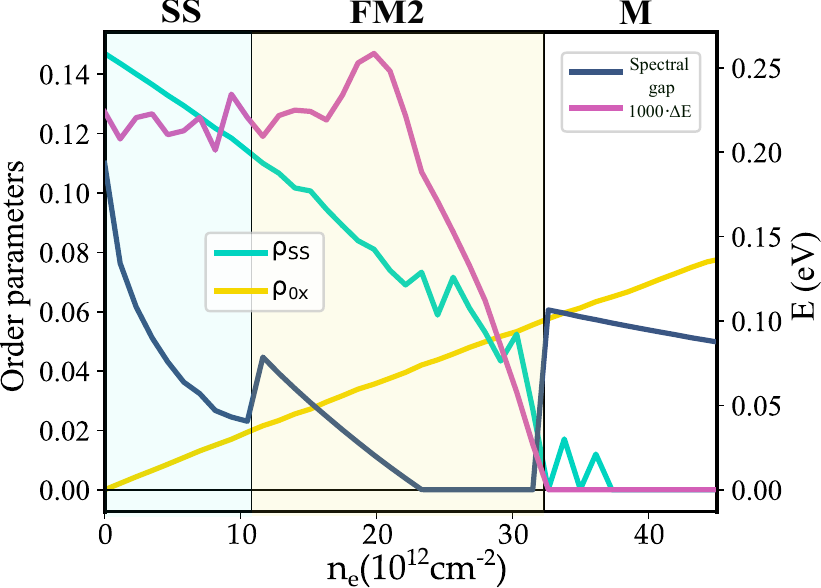}
\caption{\textbf{Sequence of ground states as a function of doping} for a fixed interaction strength $\epsilon^{-1}=0.2$.
a) The ground states are indicated by the colored shadowed area: SS (light blue), FM2 (yellow) and normal state (white). The SS-FM2 and FM2-normal state transitions are signaled by thin black lines. Light blue and yellow lines on the left vertical axis correspond to the order parameters of the SS ($\rho_{SS}$) and FM2 ($\rho_{0x}$) states. The fact that they do not vanish at the transitions indicate that these are first order. The dark blue and magenta lines on the right axis are the spectral gap at charge neutrality 
and the energy difference between the ground state and the normal metal respectively (note the latter is multiplied by a factor of 1000). The gap decreases with doping in both the SS and FM2 phases, until vanishing within the FM2 phase, and it resets to a finite value after the transition to the normal state.}
\label{fig:doping}
\end{figure}

\subsection{Evolution of the phase diagram with doping}

We extend the previous analysis to examine the competition between generalized ferromagnets and non-uniform solutions as a function of electron density leading to the complete phase diagram in Fig.~\ref{fig:full}. The SDW dome appearing at weak interactions is suppressed at a relatively small doping. 
On the contrary, the SS spreads over a wide range of $\epsilon$ values and survives until a larger critical doping, which increases with the interaction strength. 
The SS phase is surrounded by different generalized ferromagnets as a consequence of the close energy competition between them.
Like in the SS, these states -with the exception of OM3- lift the spin degeneracy of the bands, by either breaking time reversal or inversion symmetry. 

Figure~\ref{fig:doping} shows the evolution of the SS and FM2 order parameters together with the spectral gap (dark blue curve) with doping at a fixed $\epsilon^{-1}=0.2$.
As we discussed before, the trivial metal (M) is prone to a Stoner instability at small electron concentrations, thus the SS state competes directly with the FM2. 
On the one hand, the SS corresponds to an excitonic instability of the system at neutrality, whose energy gain stems from from the splitting of the valence band (see Fig.~\ref{fig:bands}(c)) as well as the increased gap with respect to the normal state. This mechanism becomes less efficient with increasing electron doping, which leads to the decrease of the SS order parameter. 
On the other hand, by increasing the doping, the FM2 order parameter and the associated spin splitting becomes larger, in such a way that the chemical potential still lies in the middle of the spin-split pockets (see Fig.~\ref{fig:bands}(b)), contrary to the case of the SS order. 
For higher dopings the order parameter continues growing, so that the exchange energy increases up to a point where the FM2 state is no longer favorable. Via this competition between kinetic and exchange energies, the FM2 always mediates between the SS at neutrality and the doped QSH state at values of $\epsilon^{-1}\leq 0.3$. These transitions are first order in our calculations. The first transition (from SS to FM2) features a reset of the spectral gap. Within the FM2 phase the spectral gap closes until it sharply opens again at the transition to the trivial metal.

At stronger interactions, $\epsilon^{-1}\geq 0.3$, the physics is different due to the undoing of the band inversion at the $\Gamma$ point in the uniform state, which makes the SS solution energetically unfavorable. At very low concentrations the transition from the SS to a trivial metal is through the OM3 solution, which is reminiscent of what happens at neutrality. As explained in Sec.~\ref{subsec:q=0}, at higher dopings, $n\geq 10^{13}$ cm$^{-2}$, the OO1/OM1 is energetically favored due to the splitting of the bands, and at even larger electron concentrations, the system evolves into the FM1 phase with the magnetization pointing along the SOC quantization axis. These transitions are marked by abrupt reductions of the spectral gap, as featured in Fig.~\ref{fig:gaps}.

\section{Discussion and Conclusions}\label{sec:conclusions}

The two leading candidates for the insulating ground state at charge neutrality and intermediate interaction strength are the QSHI and the SS. This is because, as was first pointed out in Ref.~\onlinecite{kwan_theory_2021}, the SS state involves no charge modulation in the U$_s$(1)-symmetric model, in agreement with local spectroscopy measurements, and contrary to the SDW solution, which would imply a charge modulation with half the period of the superlattice. Both the QSHI and the SS states host edge states in a finite geometry \cite{wang_breakdown_2023}, but they can be distinguished by quasiparticle interference \cite{papaj_spectroscopic_2024}.

\begin{figure}[t!] 
\includegraphics[width=\linewidth]{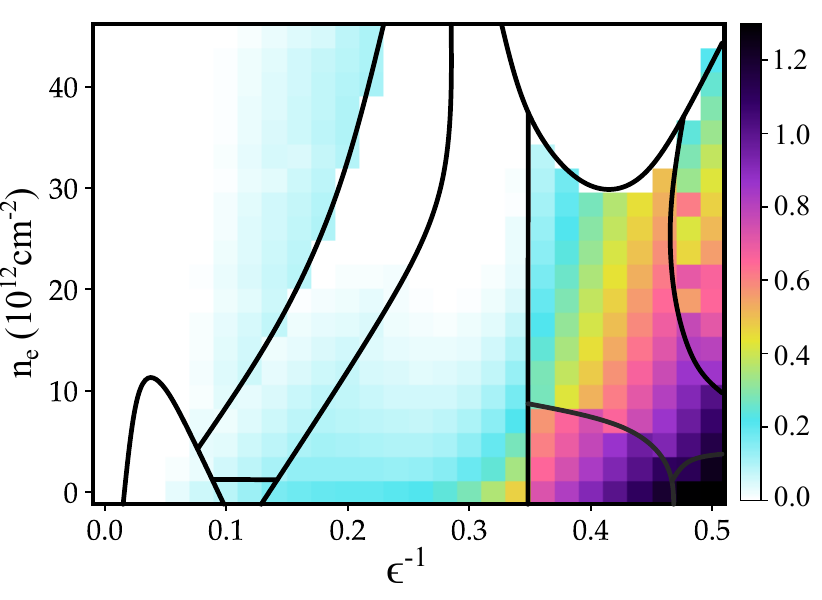}%{Fig7.pdf}
\caption{\textbf{Spectral gap in the ground state}. Black lines indicates the approximate boundaries between phases. Values of the spectral gap in the colorbar are in eV units.}
\label{fig:gaps}
\end{figure}

One of the main conclusions of our study is that upon electron doping either the QSHI or the SS state, the metal becomes an easy-plane ferromagnet (FM2) due to a Stoner instability. Adding more electrons to this ferromagnet yields a second transition to a metal consisting of the doped bands of the QSHI. In our calculations, this sequence of transitions involves resets of the spectral gap, which can be observed in tunneling or photoemission spectroscopies. 

Figure~\ref{fig:gaps} maps the spectral gap across our phase diagram deduced from our calculations. Reference~\onlinecite{que_a-gate-tunable_2024} reported a gate-driven quantum phase transition characterized by an abrupt collapse of the spectral gap. In our calculations, we find an abrupt collapse of the spectral gap only at strong interactions ($\epsilon^{-1}\geq 0.35$), by doping the OM3 state at neutrality. If, however, we start by doping the QSHI or the SS state, the changes in the spectral gap are smoother and the resets at the phase transitions are of opposite sign.
Nevertheless, our calculations probably overestimate the magnitude of the gap in all cases. 
This is because we decided not to implement a substraction scheme in the semimetallic two-band low-energy model. The reason is that if we take the semimetal as the reference state in a substraction scheme, there is no gap opening at neutrality, even if SDW or SS states are formed; or in other words: the onset of excitonic order is not enough to open a gap in the semimetal. Alternatively, one could repeat the Hartree-Fock calculations taking as the reference state the insulator reported in several density functional studies~\cite{zheng_on_2016,tang_quantum_2017,xu_2018_electrically,ok_2019_custodial,wu_quasiparticle_2024}.
Regardless of these effects, due to the physical intuition we have established for the SDW, SS, FM2 and OM3 instabilities, we expect these orders to be competitive in the phase diagram for any substraction scheme. This motivates the experimental search for magnetic phases in proximity to the spectral gap reset~\cite{que_a-gate-tunable_2024} and the superconducting phase~\cite{fatemi_electrically_2018,sajadi_gate-induced_2018,song_unconventional_2024,Song_etal2025}.

The different broken symmetries in our phase diagram may have different imprints also on the optical and transport properties of the system. 
Because time-reversal symmetry is broken in the SS state, disorder can produce backscattering between helical edge modes. 
That could explain the loss of conductance quantization as the channel length grows observed in some devices~\cite{wu_observation_2018,fei_edge_2017,tang_quantum_2017,wu_observation_2018}.
Note also that at charge neutrality, neither a sliding CDW nor a SDW can conduct electricity. However, the coherent precession of doublet solutions (in particular of the spin spiral) may support nearly dissipationless collective transport of angular momentum polarized along the spin quantization axis (as a spin \textit{superfluid} \cite{Konig_etal2001,Takei_Tserkovnyak2014}). This proposal is analogous to that for the N\'eel canted antiferromagnet in the $\nu=0$ quantum Hall state of graphene \cite{Takei_etal2016}. In the case of the OM3 state, both spatial inversion and time reversal symmetries are broken, but not their product. This state should feature a magnetoelectric response ~\cite{amaricci_exciton_2023}. Finally, signatures of broken time-reversal symmetry in the easy-plane FM2 phase may be challenging to detect. For example, the intrinsic anomalous Hall conductivity cancels exactly in the U$_s$(1)-symmetric model. The reason is that the FM2 solutions still preserves a \textit{pseudo-time reversal symmetry} resulting from the combination of $\mathcal{T}$ with a U$_s$(1) $\pi$-rotation, which gives rise to an anti-unitary operation that forces the Berry curvature to be zero (but since it squares to $+1$, the two-fold degeneracy of the spectrum is not protected, so the bands are spin-split in Fig.~\ref{fig:bands}~b). 
Another important aspect is that the energy difference between the FM2 solution and the parent fully symmetric metal is only of the order of 0.1 meV and decays very quickly with doping (see Fig.~\ref{fig:doping}), thus thermal fluctuations may destroy the FM2 order. 

In conclusion, our Hartree-Fock calculations reveal a more complex phase competition in $1T'$-WTe$_2$ monolayers than previously reported. In particular, a magnetoelectric phase with both magnetic and orbital order mediates between the reported SS excitonic and trivial insulators at charge neutrality. This phase appears at relatively high interaction strengths, roughly when the band inversion of the original model is frustrated by interactions in the parent state with no symmetry breaking. The band touching is therefore avoided by the emergence of this phase, as has been found in calculations within the Bernevig-Hughes-Zhang model with interactions~\cite{xue_time-reversal_2018,blason_exciton_2020,amaricci_exciton_2023}. In this regime we find a sequence of first-order phase transitions as a function of electron doping signaled by abrupt resets of the spectral gap. At intermediate couplings, however, the band inversion is preserved and either the QSHI or the SS state are the most likely insulating states at charge neutrality; both host helical edge modes within the gap~\cite{wang_breakdown_2023}. With electron doping these states evolve into an easy-plane ferromagnet due to a Stoner instability. Moreover, we find that this solution always appears between the insulator at neutrality and the metal consisting of the doped QSHI bands. These findings support the idea that the reported superconductivity~\cite{fatemi_electrically_2018,sajadi_gate-induced_2018,song_unconventional_2024,Song_etal2025} may arise in proximity to or from a state with broken time-reversal symmetry where both spin, intra- and inter-orbital fluctuations are potentially relevant.

\section*{Acknowledgments}
IMDEA Nanociencia acknowledges support from the ``Severo Ochoa'' Programme for Centres of Excellence in R\&D (CEX2020-001039-S/AEI/10.13039/501100011033). 
The IMDEA Nanociencia team acknowledges support from NOVMOMAT, project PID2022-142162NB-I00 funded by MICIU/AEI/10.13039/501100011033 and by FEDER, UE as well as financial support through the (MAD2D-CM)-MRR MATERIALES AVANZADOS-IMDEA-NC.
J.A. S.-G. has received financial support through the ``Ram\'on y Cajal'' Fellowship program, grant RYC2023-044383-I financed by MICIU/AEI
/10.13039/501100011033 and FSE+.
G.P.-M. is supported by Comunidad de Madrid through the PIPF2022 programme (grant number PIPF-2022TEC-26326). D.M.-S. acknowledges support from the National Science Foundation (NSF) Materials Research Science and Engineering Centers (MRSEC) program through Columbia University under the Precision-Assembled Quantum Materials (PAQM) Grant No. DMR-2011738.

\bibliography{exctonic}

\begin{thebibliography}{41}%
\makeatletter
\providecommand \@ifxundefined [1]{%
 \@ifx{#1\undefined}
}%
\providecommand \@ifnum [1]{%
 \ifnum #1\expandafter \@firstoftwo
 \else \expandafter \@secondoftwo
 \fi
}%
\providecommand \@ifx [1]{%
 \ifx #1\expandafter \@firstoftwo
 \else \expandafter \@secondoftwo
 \fi
}%
\providecommand \natexlab [1]{#1}%
\providecommand \enquote  [1]{``#1''}%
\providecommand \bibnamefont  [1]{#1}%
\providecommand \bibfnamefont [1]{#1}%
\providecommand \citenamefont [1]{#1}%
\providecommand \href@noop [0]{\@secondoftwo}%
\providecommand \href [0]{\begingroup \@sanitize@url \@href}%
\providecommand \@href[1]{\@@startlink{#1}\@@href}%
\providecommand \@@href[1]{\endgroup#1\@@endlink}%
\providecommand \@sanitize@url [0]{\catcode `\\12\catcode `\$12\catcode `\&12\catcode `\#12\catcode `\^12\catcode `\_12\catcode `\%12\relax}%
\providecommand \@@startlink[1]{}%
\providecommand \@@endlink[0]{}%
\providecommand \url  [0]{\begingroup\@sanitize@url \@url }%
\providecommand \@url [1]{\endgroup\@href {#1}{\urlprefix }}%
\providecommand \urlprefix  [0]{URL }%
\providecommand \Eprint [0]{\href }%
\providecommand \doibase [0]{https://doi.org/}%
\providecommand \selectlanguage [0]{\@gobble}%
\providecommand \bibinfo  [0]{\@secondoftwo}%
\providecommand \bibfield  [0]{\@secondoftwo}%
\providecommand \translation [1]{[#1]}%
\providecommand \BibitemOpen [0]{}%
\providecommand \bibitemStop [0]{}%
\providecommand \bibitemNoStop [0]{.\EOS\space}%
\providecommand \EOS [0]{\spacefactor3000\relax}%
\providecommand \BibitemShut  [1]{\csname bibitem#1\endcsname}%
\let\auto@bib@innerbib\@empty
%</preamble>
\bibitem [{\citenamefont {Rold{\'a}n}\ \emph {et~al.}(2017)\citenamefont {Rold{\'a}n}, \citenamefont {Chirolli}, \citenamefont {Prada}, \citenamefont {Silva-Guill{\'e}n}, \citenamefont {San-Jose},\ and\ \citenamefont {Guinea}}]{roldan_theory_2017}%
  \BibitemOpen
  \bibfield  {author} {\bibinfo {author} {\bibfnamefont {R.}~\bibnamefont {Rold{\'a}n}}, \bibinfo {author} {\bibfnamefont {L.}~\bibnamefont {Chirolli}}, \bibinfo {author} {\bibfnamefont {E.}~\bibnamefont {Prada}}, \bibinfo {author} {\bibfnamefont {J.~A.}\ \bibnamefont {Silva-Guill{\'e}n}}, \bibinfo {author} {\bibfnamefont {P.}~\bibnamefont {San-Jose}},\ and\ \bibinfo {author} {\bibfnamefont {F.}~\bibnamefont {Guinea}},\ }\bibfield  {title} {\bibinfo {title} {Theory of {2D} crystals: {Graphene} and beyond},\ }\href {https://doi.org/10.1039/c7cs00210f} {\bibfield  {journal} {\bibinfo  {journal} {Chemical Society Reviews}\ }\textbf {\bibinfo {volume} {46}},\ \bibinfo {pages} {4387} (\bibinfo {year} {2017})}\BibitemShut {NoStop}%
\bibitem [{\citenamefont {Manzeli}\ \emph {et~al.}(2017)\citenamefont {Manzeli}, \citenamefont {Ovchinnikov}, \citenamefont {Pasquier}, \citenamefont {Yazyev},\ and\ \citenamefont {Kis}}]{manzeli_2d_2017}%
  \BibitemOpen
  \bibfield  {author} {\bibinfo {author} {\bibfnamefont {S.}~\bibnamefont {Manzeli}}, \bibinfo {author} {\bibfnamefont {D.}~\bibnamefont {Ovchinnikov}}, \bibinfo {author} {\bibfnamefont {D.}~\bibnamefont {Pasquier}}, \bibinfo {author} {\bibfnamefont {O.~V.}\ \bibnamefont {Yazyev}},\ and\ \bibinfo {author} {\bibfnamefont {A.}~\bibnamefont {Kis}},\ }\bibfield  {title} {\bibinfo {title} {{2}{D} transition metal dichalcogenides},\ }\href {https://doi.org/10.1038/natrevmats.2017.33} {\bibfield  {journal} {\bibinfo  {journal} {Nature Reviews Materials}\ }\textbf {\bibinfo {volume} {2}},\ \bibinfo {pages} {1} (\bibinfo {year} {2017})}\BibitemShut {NoStop}%
\bibitem [{\citenamefont {Choi}\ \emph {et~al.}(2017)\citenamefont {Choi}, \citenamefont {Choudhary}, \citenamefont {Han}, \citenamefont {Park}, \citenamefont {Akinwande},\ and\ \citenamefont {Lee}}]{choi_2017_recent}%
  \BibitemOpen
  \bibfield  {author} {\bibinfo {author} {\bibfnamefont {W.}~\bibnamefont {Choi}}, \bibinfo {author} {\bibfnamefont {N.}~\bibnamefont {Choudhary}}, \bibinfo {author} {\bibfnamefont {G.~H.}\ \bibnamefont {Han}}, \bibinfo {author} {\bibfnamefont {J.}~\bibnamefont {Park}}, \bibinfo {author} {\bibfnamefont {D.}~\bibnamefont {Akinwande}},\ and\ \bibinfo {author} {\bibfnamefont {Y.~H.}\ \bibnamefont {Lee}},\ }\bibfield  {title} {\bibinfo {title} {Recent development of two-dimensional transition metal dichalcogenides and their applications},\ }\href {https://doi.org/https://doi.org/10.1016/j.mattod.2016.10.002} {\bibfield  {journal} {\bibinfo  {journal} {Materials Today}\ }\textbf {\bibinfo {volume} {20}},\ \bibinfo {pages} {116} (\bibinfo {year} {2017})}\BibitemShut {NoStop}%
\bibitem [{\citenamefont {Huang}\ \emph {et~al.}(2020)\citenamefont {Huang}, \citenamefont {Fan}, \citenamefont {Singh},\ and\ \citenamefont {Zheng}}]{huang_2020_recent}%
  \BibitemOpen
  \bibfield  {author} {\bibinfo {author} {\bibfnamefont {H.~H.}\ \bibnamefont {Huang}}, \bibinfo {author} {\bibfnamefont {X.}~\bibnamefont {Fan}}, \bibinfo {author} {\bibfnamefont {D.~J.}\ \bibnamefont {Singh}},\ and\ \bibinfo {author} {\bibfnamefont {W.~T.}\ \bibnamefont {Zheng}},\ }\bibfield  {title} {\bibinfo {title} {Recent progress of tmd nanomaterials: phase transitions and applications},\ }\href {https://doi.org/10.1039/C9NR08313H} {\bibfield  {journal} {\bibinfo  {journal} {Nanoscale}\ }\textbf {\bibinfo {volume} {12}},\ \bibinfo {pages} {1247} (\bibinfo {year} {2020})}\BibitemShut {NoStop}%
\bibitem [{\citenamefont {Siddique}\ \emph {et~al.}(2021)\citenamefont {Siddique}, \citenamefont {Gowda}, \citenamefont {Demiss}, \citenamefont {Tromer}, \citenamefont {Paul}, \citenamefont {Sadasivuni}, \citenamefont {Olu}, \citenamefont {Chandra}, \citenamefont {Kochat}, \citenamefont {Galv{\~a}o}, \citenamefont {Kumbhakar}, \citenamefont {Mishra}, \citenamefont {Aayan},\ and\ \citenamefont {Tiwary}}]{siddique_2021_emerging}%
  \BibitemOpen
  \bibfield  {author} {\bibinfo {author} {\bibfnamefont {S.}~\bibnamefont {Siddique}}, \bibinfo {author} {\bibfnamefont {C.~C.}\ \bibnamefont {Gowda}}, \bibinfo {author} {\bibfnamefont {S.}~\bibnamefont {Demiss}}, \bibinfo {author} {\bibfnamefont {R.}~\bibnamefont {Tromer}}, \bibinfo {author} {\bibfnamefont {S.}~\bibnamefont {Paul}}, \bibinfo {author} {\bibfnamefont {K.~K.}\ \bibnamefont {Sadasivuni}}, \bibinfo {author} {\bibfnamefont {E.~F.}\ \bibnamefont {Olu}}, \bibinfo {author} {\bibfnamefont {A.}~\bibnamefont {Chandra}}, \bibinfo {author} {\bibfnamefont {V.}~\bibnamefont {Kochat}}, \bibinfo {author} {\bibfnamefont {D.~S.}\ \bibnamefont {Galv{\~a}o}}, \bibinfo {author} {\bibfnamefont {P.}~\bibnamefont {Kumbhakar}}, \bibinfo {author} {\bibfnamefont {R.}~\bibnamefont {Mishra}}, \bibinfo {author} {\bibfnamefont {P.~M.}\ \bibnamefont {Aayan}},\ and\ \bibinfo {author} {\bibfnamefont {C.~S.}\ \bibnamefont {Tiwary}},\ }\bibfield  {title} {\bibinfo {title} {Emerging two-dimensional tellurides},\ }\href
  {https://doi.org/https://doi.org/10.1016/j.mattod.2021.08.008} {\bibfield  {journal} {\bibinfo  {journal} {Materials Today}\ }\textbf {\bibinfo {volume} {51}},\ \bibinfo {pages} {402} (\bibinfo {year} {2021})}\BibitemShut {NoStop}%
\bibitem [{\citenamefont {Roldan}\ \emph {et~al.}(2014)\citenamefont {Roldan}, \citenamefont {Silva-Guillen}, \citenamefont {Lopez-Sancho}, \citenamefont {Guinea}, \citenamefont {Cappelluti},\ and\ \citenamefont {Ordejon}}]{roldan_electronic_2014}%
  \BibitemOpen
  \bibfield  {author} {\bibinfo {author} {\bibfnamefont {R.}~\bibnamefont {Roldan}}, \bibinfo {author} {\bibfnamefont {J.~A.}\ \bibnamefont {Silva-Guillen}}, \bibinfo {author} {\bibfnamefont {M.~P.}\ \bibnamefont {Lopez-Sancho}}, \bibinfo {author} {\bibfnamefont {F.}~\bibnamefont {Guinea}}, \bibinfo {author} {\bibfnamefont {E.}~\bibnamefont {Cappelluti}},\ and\ \bibinfo {author} {\bibfnamefont {P.}~\bibnamefont {Ordejon}},\ }\bibfield  {title} {\bibinfo {title} {Electronic properties of single-layer and multilayer transition metal dichalcogenides {$M$}{$X$}$_2$ ({$M$} = {Mo}, {W} and {$X$} = {S}, {Se})},\ }\href {https://doi.org/10.1002/andp.201400128} {\bibfield  {journal} {\bibinfo  {journal} {Annalen Der Physik}\ }\textbf {\bibinfo {volume} {526}},\ \bibinfo {pages} {347} (\bibinfo {year} {2014})}\BibitemShut {NoStop}%
\bibitem [{\citenamefont {Fei}\ \emph {et~al.}(2017)\citenamefont {Fei}, \citenamefont {Palomaki}, \citenamefont {Wu}, \citenamefont {Zhao}, \citenamefont {Cai}, \citenamefont {Sun}, \citenamefont {Nguyen}, \citenamefont {Finney}, \citenamefont {Xu},\ and\ \citenamefont {Cobden}}]{fei_edge_2017}%
  \BibitemOpen
  \bibfield  {author} {\bibinfo {author} {\bibfnamefont {Z.}~\bibnamefont {Fei}}, \bibinfo {author} {\bibfnamefont {T.}~\bibnamefont {Palomaki}}, \bibinfo {author} {\bibfnamefont {S.}~\bibnamefont {Wu}}, \bibinfo {author} {\bibfnamefont {W.}~\bibnamefont {Zhao}}, \bibinfo {author} {\bibfnamefont {X.}~\bibnamefont {Cai}}, \bibinfo {author} {\bibfnamefont {B.}~\bibnamefont {Sun}}, \bibinfo {author} {\bibfnamefont {P.}~\bibnamefont {Nguyen}}, \bibinfo {author} {\bibfnamefont {J.}~\bibnamefont {Finney}}, \bibinfo {author} {\bibfnamefont {X.}~\bibnamefont {Xu}},\ and\ \bibinfo {author} {\bibfnamefont {D.~H.}\ \bibnamefont {Cobden}},\ }\bibfield  {title} {\bibinfo {title} {Edge conduction in monolayer {W}{Te}{$_2$}},\ }\href {https://doi.org/10.1038/nphys4091} {\bibfield  {journal} {\bibinfo  {journal} {Nature Physics}\ }\textbf {\bibinfo {volume} {13}},\ \bibinfo {pages} {677} (\bibinfo {year} {2017})}\BibitemShut {NoStop}%
\bibitem [{\citenamefont {Tang}\ \emph {et~al.}(2017)\citenamefont {Tang}, \citenamefont {Zhang}, \citenamefont {Wong}, \citenamefont {Pedramrazi}, \citenamefont {Tsai}, \citenamefont {Jia}, \citenamefont {Moritz}, \citenamefont {Claassen}, \citenamefont {Ryu}, \citenamefont {Kahn}, \citenamefont {Jiang}, \citenamefont {Yan}, \citenamefont {Hashimoto}, \citenamefont {Lu}, \citenamefont {Moore}, \citenamefont {Hwang}, \citenamefont {Hwang}, \citenamefont {Hussain}, \citenamefont {Chen}, \citenamefont {Ugeda}, \citenamefont {Liu}, \citenamefont {Xie}, \citenamefont {Devereaux}, \citenamefont {Crommie}, \citenamefont {Mo},\ and\ \citenamefont {Shen}}]{tang_quantum_2017}%
  \BibitemOpen
  \bibfield  {author} {\bibinfo {author} {\bibfnamefont {S.}~\bibnamefont {Tang}}, \bibinfo {author} {\bibfnamefont {C.}~\bibnamefont {Zhang}}, \bibinfo {author} {\bibfnamefont {D.}~\bibnamefont {Wong}}, \bibinfo {author} {\bibfnamefont {Z.}~\bibnamefont {Pedramrazi}}, \bibinfo {author} {\bibfnamefont {H.-Z.}\ \bibnamefont {Tsai}}, \bibinfo {author} {\bibfnamefont {C.}~\bibnamefont {Jia}}, \bibinfo {author} {\bibfnamefont {B.}~\bibnamefont {Moritz}}, \bibinfo {author} {\bibfnamefont {M.}~\bibnamefont {Claassen}}, \bibinfo {author} {\bibfnamefont {H.}~\bibnamefont {Ryu}}, \bibinfo {author} {\bibfnamefont {S.}~\bibnamefont {Kahn}}, \bibinfo {author} {\bibfnamefont {J.}~\bibnamefont {Jiang}}, \bibinfo {author} {\bibfnamefont {H.}~\bibnamefont {Yan}}, \bibinfo {author} {\bibfnamefont {M.}~\bibnamefont {Hashimoto}}, \bibinfo {author} {\bibfnamefont {D.}~\bibnamefont {Lu}}, \bibinfo {author} {\bibfnamefont {R.~G.}\ \bibnamefont {Moore}}, \bibinfo {author} {\bibfnamefont {C.-C.}\ \bibnamefont {Hwang}}, \bibinfo
  {author} {\bibfnamefont {C.}~\bibnamefont {Hwang}}, \bibinfo {author} {\bibfnamefont {Z.}~\bibnamefont {Hussain}}, \bibinfo {author} {\bibfnamefont {Y.}~\bibnamefont {Chen}}, \bibinfo {author} {\bibfnamefont {M.~M.}\ \bibnamefont {Ugeda}}, \bibinfo {author} {\bibfnamefont {Z.}~\bibnamefont {Liu}}, \bibinfo {author} {\bibfnamefont {X.}~\bibnamefont {Xie}}, \bibinfo {author} {\bibfnamefont {T.~P.}\ \bibnamefont {Devereaux}}, \bibinfo {author} {\bibfnamefont {M.~F.}\ \bibnamefont {Crommie}}, \bibinfo {author} {\bibfnamefont {S.-K.}\ \bibnamefont {Mo}},\ and\ \bibinfo {author} {\bibfnamefont {Z.-X.}\ \bibnamefont {Shen}},\ }\bibfield  {title} {\bibinfo {title} {Quantum spin {Hall} state in monolayer {1T}'-{W}{Te}{$_2$}},\ }\href {https://doi.org/10.1038/nphys4174} {\bibfield  {journal} {\bibinfo  {journal} {Nature Physics}\ }\textbf {\bibinfo {volume} {13}},\ \bibinfo {pages} {683} (\bibinfo {year} {2017})}\BibitemShut {NoStop}%
\bibitem [{\citenamefont {Wu}\ \emph {et~al.}(2018)\citenamefont {Wu}, \citenamefont {Fatemi}, \citenamefont {Gibson}, \citenamefont {Watanabe}, \citenamefont {Taniguchi}, \citenamefont {Cava},\ and\ \citenamefont {Jarillo-Herrero}}]{wu_observation_2018}%
  \BibitemOpen
  \bibfield  {author} {\bibinfo {author} {\bibfnamefont {S.}~\bibnamefont {Wu}}, \bibinfo {author} {\bibfnamefont {V.}~\bibnamefont {Fatemi}}, \bibinfo {author} {\bibfnamefont {Q.~D.}\ \bibnamefont {Gibson}}, \bibinfo {author} {\bibfnamefont {K.}~\bibnamefont {Watanabe}}, \bibinfo {author} {\bibfnamefont {T.}~\bibnamefont {Taniguchi}}, \bibinfo {author} {\bibfnamefont {R.~J.}\ \bibnamefont {Cava}},\ and\ \bibinfo {author} {\bibfnamefont {P.}~\bibnamefont {Jarillo-Herrero}},\ }\bibfield  {title} {\bibinfo {title} {Observation of the quantum spin {H}all effect up to 100 kelvin in a monolayer crystal},\ }\href {https://doi.org/10.1126/science.aan6003} {\bibfield  {journal} {\bibinfo  {journal} {Science}\ }\textbf {\bibinfo {volume} {359}},\ \bibinfo {pages} {76} (\bibinfo {year} {2018})}\BibitemShut {NoStop}%
\bibitem [{\citenamefont {Fatemi}\ \emph {et~al.}(2018)\citenamefont {Fatemi}, \citenamefont {Wu}, \citenamefont {Cao}, \citenamefont {Bretheau}, \citenamefont {Gibson}, \citenamefont {Watanabe}, \citenamefont {Taniguchi}, \citenamefont {Cava},\ and\ \citenamefont {Jarillo-Herrero}}]{fatemi_electrically_2018}%
  \BibitemOpen
  \bibfield  {author} {\bibinfo {author} {\bibfnamefont {V.}~\bibnamefont {Fatemi}}, \bibinfo {author} {\bibfnamefont {S.}~\bibnamefont {Wu}}, \bibinfo {author} {\bibfnamefont {Y.}~\bibnamefont {Cao}}, \bibinfo {author} {\bibfnamefont {L.}~\bibnamefont {Bretheau}}, \bibinfo {author} {\bibfnamefont {Q.~D.}\ \bibnamefont {Gibson}}, \bibinfo {author} {\bibfnamefont {K.}~\bibnamefont {Watanabe}}, \bibinfo {author} {\bibfnamefont {T.}~\bibnamefont {Taniguchi}}, \bibinfo {author} {\bibfnamefont {R.~J.}\ \bibnamefont {Cava}},\ and\ \bibinfo {author} {\bibfnamefont {P.}~\bibnamefont {Jarillo-Herrero}},\ }\bibfield  {title} {\bibinfo {title} {Electrically tunable low-density superconductivity in a monolayer topological insulator},\ }\href {https://doi.org/10.1126/science.aar4642} {\bibfield  {journal} {\bibinfo  {journal} {Science}\ }\textbf {\bibinfo {volume} {362}},\ \bibinfo {pages} {926} (\bibinfo {year} {2018})}\BibitemShut {NoStop}%
\bibitem [{\citenamefont {Sajadi}\ \emph {et~al.}(2018)\citenamefont {Sajadi}, \citenamefont {Palomaki}, \citenamefont {Fei}, \citenamefont {Zhao}, \citenamefont {Bement}, \citenamefont {Olsen}, \citenamefont {Luescher}, \citenamefont {Xu}, \citenamefont {Folk},\ and\ \citenamefont {Cobden}}]{sajadi_gate-induced_2018}%
  \BibitemOpen
  \bibfield  {author} {\bibinfo {author} {\bibfnamefont {E.}~\bibnamefont {Sajadi}}, \bibinfo {author} {\bibfnamefont {T.}~\bibnamefont {Palomaki}}, \bibinfo {author} {\bibfnamefont {Z.}~\bibnamefont {Fei}}, \bibinfo {author} {\bibfnamefont {W.}~\bibnamefont {Zhao}}, \bibinfo {author} {\bibfnamefont {P.}~\bibnamefont {Bement}}, \bibinfo {author} {\bibfnamefont {C.}~\bibnamefont {Olsen}}, \bibinfo {author} {\bibfnamefont {S.}~\bibnamefont {Luescher}}, \bibinfo {author} {\bibfnamefont {X.}~\bibnamefont {Xu}}, \bibinfo {author} {\bibfnamefont {J.~A.}\ \bibnamefont {Folk}},\ and\ \bibinfo {author} {\bibfnamefont {D.~H.}\ \bibnamefont {Cobden}},\ }\bibfield  {title} {\bibinfo {title} {Gate-induced superconductivity in a monolayer topological insulator},\ }\href {https://doi.org/10.1126/science.aar4426} {\bibfield  {journal} {\bibinfo  {journal} {Science}\ }\textbf {\bibinfo {volume} {362}},\ \bibinfo {pages} {922} (\bibinfo {year} {2018})}\BibitemShut {NoStop}%
\bibitem [{\citenamefont {Song}\ \emph {et~al.}(2024)\citenamefont {Song}, \citenamefont {Jia}, \citenamefont {Yu}, \citenamefont {Tang}, \citenamefont {Wang}, \citenamefont {Singha}, \citenamefont {Gui}, \citenamefont {Uzan-Narovlansky}, \citenamefont {Onyszczak}, \citenamefont {Watanabe}, \citenamefont {Taniguchi}, \citenamefont {Cava}, \citenamefont {Schoop}, \citenamefont {Ong},\ and\ \citenamefont {Wu}}]{song_unconventional_2024}%
  \BibitemOpen
  \bibfield  {author} {\bibinfo {author} {\bibfnamefont {T.}~\bibnamefont {Song}}, \bibinfo {author} {\bibfnamefont {Y.}~\bibnamefont {Jia}}, \bibinfo {author} {\bibfnamefont {G.}~\bibnamefont {Yu}}, \bibinfo {author} {\bibfnamefont {Y.}~\bibnamefont {Tang}}, \bibinfo {author} {\bibfnamefont {P.}~\bibnamefont {Wang}}, \bibinfo {author} {\bibfnamefont {R.}~\bibnamefont {Singha}}, \bibinfo {author} {\bibfnamefont {X.}~\bibnamefont {Gui}}, \bibinfo {author} {\bibfnamefont {A.~J.}\ \bibnamefont {Uzan-Narovlansky}}, \bibinfo {author} {\bibfnamefont {M.}~\bibnamefont {Onyszczak}}, \bibinfo {author} {\bibfnamefont {K.}~\bibnamefont {Watanabe}}, \bibinfo {author} {\bibfnamefont {T.}~\bibnamefont {Taniguchi}}, \bibinfo {author} {\bibfnamefont {R.~J.}\ \bibnamefont {Cava}}, \bibinfo {author} {\bibfnamefont {L.~M.}\ \bibnamefont {Schoop}}, \bibinfo {author} {\bibfnamefont {N.~P.}\ \bibnamefont {Ong}},\ and\ \bibinfo {author} {\bibfnamefont {S.}~\bibnamefont {Wu}},\ }\bibfield  {title} {\bibinfo {title} {Unconventional
  superconducting quantum criticality in monolayer {W}{Te}{$_2$}},\ }\href {https://doi.org/10.1038/s41567-023-02291-1} {\bibfield  {journal} {\bibinfo  {journal} {Nature Physics}\ }\textbf {\bibinfo {volume} {20}},\ \bibinfo {pages} {269} (\bibinfo {year} {2024})}\BibitemShut {NoStop}%
\bibitem [{\citenamefont {Song}\ \emph {et~al.}(2025)\citenamefont {Song}, \citenamefont {Jia}, \citenamefont {Yu}, \citenamefont {Tang}, \citenamefont {Uzan}, \citenamefont {Zheng}, \citenamefont {Guan}, \citenamefont {Onyszczak}, \citenamefont {Singha}, \citenamefont {Gui}, \citenamefont {Watanabe}, \citenamefont {Taniguchi}, \citenamefont {Cava}, \citenamefont {Schoop}, \citenamefont {Ong},\ and\ \citenamefont {Wu}}]{Song_etal2025}%
  \BibitemOpen
  \bibfield  {author} {\bibinfo {author} {\bibfnamefont {T.}~\bibnamefont {Song}}, \bibinfo {author} {\bibfnamefont {Y.}~\bibnamefont {Jia}}, \bibinfo {author} {\bibfnamefont {G.}~\bibnamefont {Yu}}, \bibinfo {author} {\bibfnamefont {Y.}~\bibnamefont {Tang}}, \bibinfo {author} {\bibfnamefont {A.~J.}\ \bibnamefont {Uzan}}, \bibinfo {author} {\bibfnamefont {Z.~J.}\ \bibnamefont {Zheng}}, \bibinfo {author} {\bibfnamefont {H.}~\bibnamefont {Guan}}, \bibinfo {author} {\bibfnamefont {M.}~\bibnamefont {Onyszczak}}, \bibinfo {author} {\bibfnamefont {R.}~\bibnamefont {Singha}}, \bibinfo {author} {\bibfnamefont {X.}~\bibnamefont {Gui}}, \bibinfo {author} {\bibfnamefont {K.}~\bibnamefont {Watanabe}}, \bibinfo {author} {\bibfnamefont {T.}~\bibnamefont {Taniguchi}}, \bibinfo {author} {\bibfnamefont {R.~J.}\ \bibnamefont {Cava}}, \bibinfo {author} {\bibfnamefont {L.~M.}\ \bibnamefont {Schoop}}, \bibinfo {author} {\bibfnamefont {N.~P.}\ \bibnamefont {Ong}},\ and\ \bibinfo {author} {\bibfnamefont {S.}~\bibnamefont {Wu}},\
  }\bibfield  {title} {\bibinfo {title} {Unconventional superconducting phase diagram of monolayer wte$_2$},\ }\href@noop {} {\bibfield  {journal} {\bibinfo  {journal} {arXiv:2501.16699}\ } (\bibinfo {year} {2025})}\BibitemShut {NoStop}%
\bibitem [{\citenamefont {Qian}\ \emph {et~al.}(2014)\citenamefont {Qian}, \citenamefont {Liu}, \citenamefont {Fu},\ and\ \citenamefont {Li}}]{qian_quantum_2014}%
  \BibitemOpen
  \bibfield  {author} {\bibinfo {author} {\bibfnamefont {X.}~\bibnamefont {Qian}}, \bibinfo {author} {\bibfnamefont {J.}~\bibnamefont {Liu}}, \bibinfo {author} {\bibfnamefont {L.}~\bibnamefont {Fu}},\ and\ \bibinfo {author} {\bibfnamefont {J.}~\bibnamefont {Li}},\ }\bibfield  {title} {\bibinfo {title} {Quantum spin {Hall} effect in two-dimensional transition metal dichalcogenides},\ }\href {https://doi.org/10.1126/science.1256815} {\bibfield  {journal} {\bibinfo  {journal} {Science}\ }\textbf {\bibinfo {volume} {346}},\ \bibinfo {pages} {1344} (\bibinfo {year} {2014})}\BibitemShut {NoStop}%
\bibitem [{\citenamefont {Zheng}\ \emph {et~al.}(2016)\citenamefont {Zheng}, \citenamefont {Cai}, \citenamefont {Ge}, \citenamefont {Zhang}, \citenamefont {Liu}, \citenamefont {Lu}, \citenamefont {Zhang}, \citenamefont {Qiu}, \citenamefont {Taniguchi}, \citenamefont {Watanabe}, \citenamefont {Jia}, \citenamefont {Qi}, \citenamefont {Chen}, \citenamefont {Sun},\ and\ \citenamefont {Feng}}]{zheng_on_2016}%
  \BibitemOpen
  \bibfield  {author} {\bibinfo {author} {\bibfnamefont {F.}~\bibnamefont {Zheng}}, \bibinfo {author} {\bibfnamefont {C.}~\bibnamefont {Cai}}, \bibinfo {author} {\bibfnamefont {S.}~\bibnamefont {Ge}}, \bibinfo {author} {\bibfnamefont {X.}~\bibnamefont {Zhang}}, \bibinfo {author} {\bibfnamefont {X.}~\bibnamefont {Liu}}, \bibinfo {author} {\bibfnamefont {H.}~\bibnamefont {Lu}}, \bibinfo {author} {\bibfnamefont {Y.}~\bibnamefont {Zhang}}, \bibinfo {author} {\bibfnamefont {J.}~\bibnamefont {Qiu}}, \bibinfo {author} {\bibfnamefont {T.}~\bibnamefont {Taniguchi}}, \bibinfo {author} {\bibfnamefont {K.}~\bibnamefont {Watanabe}}, \bibinfo {author} {\bibfnamefont {S.}~\bibnamefont {Jia}}, \bibinfo {author} {\bibfnamefont {J.}~\bibnamefont {Qi}}, \bibinfo {author} {\bibfnamefont {J.-H.}\ \bibnamefont {Chen}}, \bibinfo {author} {\bibfnamefont {D.}~\bibnamefont {Sun}},\ and\ \bibinfo {author} {\bibfnamefont {J.}~\bibnamefont {Feng}},\ }\bibfield  {title} {\bibinfo {title} {On the quantum spin hall gap of monolayer
  {1T}'-wte2},\ }\href {https://doi.org/https://doi.org/10.1002/adma.201600100} {\bibfield  {journal} {\bibinfo  {journal} {Advanced Materials}\ }\textbf {\bibinfo {volume} {28}},\ \bibinfo {pages} {4845} (\bibinfo {year} {2016})}\BibitemShut {NoStop}%
\bibitem [{\citenamefont {Xu}\ \emph {et~al.}(2018)\citenamefont {Xu}, \citenamefont {Ma}, \citenamefont {Shen}, \citenamefont {Fatemi}, \citenamefont {Wu}, \citenamefont {Chang}, \citenamefont {Chang}, \citenamefont {Valdivia}, \citenamefont {Chan}, \citenamefont {Gibson} \emph {et~al.}}]{xu_2018_electrically}%
  \BibitemOpen
  \bibfield  {author} {\bibinfo {author} {\bibfnamefont {S.-Y.}\ \bibnamefont {Xu}}, \bibinfo {author} {\bibfnamefont {Q.}~\bibnamefont {Ma}}, \bibinfo {author} {\bibfnamefont {H.}~\bibnamefont {Shen}}, \bibinfo {author} {\bibfnamefont {V.}~\bibnamefont {Fatemi}}, \bibinfo {author} {\bibfnamefont {S.}~\bibnamefont {Wu}}, \bibinfo {author} {\bibfnamefont {T.-R.}\ \bibnamefont {Chang}}, \bibinfo {author} {\bibfnamefont {G.}~\bibnamefont {Chang}}, \bibinfo {author} {\bibfnamefont {A.~M.~M.}\ \bibnamefont {Valdivia}}, \bibinfo {author} {\bibfnamefont {C.-K.}\ \bibnamefont {Chan}}, \bibinfo {author} {\bibfnamefont {Q.~D.}\ \bibnamefont {Gibson}}, \emph {et~al.},\ }\bibfield  {title} {\bibinfo {title} {Electrically switchable berry curvature dipole in the monolayer topological insulator wte2},\ }\href {https://doi.org/10.1038/s41567-018-0189-6} {\bibfield  {journal} {\bibinfo  {journal} {Nature Physics}\ }\textbf {\bibinfo {volume} {14}},\ \bibinfo {pages} {900} (\bibinfo {year} {2018})}\BibitemShut {NoStop}%
\bibitem [{\citenamefont {Ok}\ \emph {et~al.}(2019)\citenamefont {Ok}, \citenamefont {Muechler}, \citenamefont {Di~Sante}, \citenamefont {Sangiovanni}, \citenamefont {Thomale},\ and\ \citenamefont {Neupert}}]{ok_2019_custodial}%
  \BibitemOpen
  \bibfield  {author} {\bibinfo {author} {\bibfnamefont {S.}~\bibnamefont {Ok}}, \bibinfo {author} {\bibfnamefont {L.}~\bibnamefont {Muechler}}, \bibinfo {author} {\bibfnamefont {D.}~\bibnamefont {Di~Sante}}, \bibinfo {author} {\bibfnamefont {G.}~\bibnamefont {Sangiovanni}}, \bibinfo {author} {\bibfnamefont {R.}~\bibnamefont {Thomale}},\ and\ \bibinfo {author} {\bibfnamefont {T.}~\bibnamefont {Neupert}},\ }\bibfield  {title} {\bibinfo {title} {Custodial glide symmetry of quantum spin hall edge modes in monolayer ${\mathrm{wte}}_{2}$},\ }\href {https://doi.org/10.1103/PhysRevB.99.121105} {\bibfield  {journal} {\bibinfo  {journal} {Phys. Rev. B}\ }\textbf {\bibinfo {volume} {99}},\ \bibinfo {pages} {121105} (\bibinfo {year} {2019})}\BibitemShut {NoStop}%
\bibitem [{\citenamefont {Wu}\ \emph {et~al.}(2024)\citenamefont {Wu}, \citenamefont {Hou}, \citenamefont {Li}, \citenamefont {He},\ and\ \citenamefont {Qiu}}]{wu_quasiparticle_2024}%
  \BibitemOpen
  \bibfield  {author} {\bibinfo {author} {\bibfnamefont {J.}~\bibnamefont {Wu}}, \bibinfo {author} {\bibfnamefont {B.}~\bibnamefont {Hou}}, \bibinfo {author} {\bibfnamefont {W.}~\bibnamefont {Li}}, \bibinfo {author} {\bibfnamefont {Y.}~\bibnamefont {He}},\ and\ \bibinfo {author} {\bibfnamefont {D.~Y.}\ \bibnamefont {Qiu}},\ }\bibfield  {title} {\bibinfo {title} {Quasiparticle and excitonic properties of monolayer 1{$T$}'-{W}{Te}{$_2$} within many-body perturbation theory},\ }\href {https://doi.org/10.1103/PhysRevB.110.075133} {\bibfield  {journal} {\bibinfo  {journal} {Physical Review B}\ }\textbf {\bibinfo {volume} {110}},\ \bibinfo {pages} {075133} (\bibinfo {year} {2024})}\BibitemShut {NoStop}%
\bibitem [{\citenamefont {Song}\ \emph {et~al.}(2018)\citenamefont {Song}, \citenamefont {Jia}, \citenamefont {Zhang}, \citenamefont {Zhu}, \citenamefont {Shi}, \citenamefont {Wang}, \citenamefont {Zhu}, \citenamefont {Yuan}, \citenamefont {Zhang}, \citenamefont {Xing} \emph {et~al.}}]{song_observation_2018}%
  \BibitemOpen
  \bibfield  {author} {\bibinfo {author} {\bibfnamefont {Y.-H.}\ \bibnamefont {Song}}, \bibinfo {author} {\bibfnamefont {Z.-Y.}\ \bibnamefont {Jia}}, \bibinfo {author} {\bibfnamefont {D.}~\bibnamefont {Zhang}}, \bibinfo {author} {\bibfnamefont {X.-Y.}\ \bibnamefont {Zhu}}, \bibinfo {author} {\bibfnamefont {Z.-Q.}\ \bibnamefont {Shi}}, \bibinfo {author} {\bibfnamefont {H.}~\bibnamefont {Wang}}, \bibinfo {author} {\bibfnamefont {L.}~\bibnamefont {Zhu}}, \bibinfo {author} {\bibfnamefont {Q.-Q.}\ \bibnamefont {Yuan}}, \bibinfo {author} {\bibfnamefont {H.}~\bibnamefont {Zhang}}, \bibinfo {author} {\bibfnamefont {D.-Y.}\ \bibnamefont {Xing}}, \emph {et~al.},\ }\bibfield  {title} {\bibinfo {title} {Observation of {C}oulomb gap in the quantum spin {H}all candidate single-layer {1T}'-{W}{Te}$_2$},\ }\href {https://doi.org/10.1038/s41467-018-06635-x} {\bibfield  {journal} {\bibinfo  {journal} {Nature Communications}\ }\textbf {\bibinfo {volume} {9}},\ \bibinfo {pages} {4071} (\bibinfo {year} {2018})}\BibitemShut
  {NoStop}%
\bibitem [{\citenamefont {Kohn}(1967)}]{kohn_excitonic_1967}%
  \BibitemOpen
  \bibfield  {author} {\bibinfo {author} {\bibfnamefont {W.}~\bibnamefont {Kohn}},\ }\bibfield  {title} {\bibinfo {title} {Excitonic {Phases}},\ }\href {https://doi.org/10.1103/PhysRevLett.19.439} {\bibfield  {journal} {\bibinfo  {journal} {Physical Review Letters}\ }\textbf {\bibinfo {volume} {19}},\ \bibinfo {pages} {439} (\bibinfo {year} {1967})}\BibitemShut {NoStop}%
\bibitem [{\citenamefont {Halperin}\ and\ \citenamefont {Rice}(1968)}]{halperin_possible_1968}%
  \BibitemOpen
  \bibfield  {author} {\bibinfo {author} {\bibfnamefont {B.~I.}\ \bibnamefont {Halperin}}\ and\ \bibinfo {author} {\bibfnamefont {T.~M.}\ \bibnamefont {Rice}},\ }\bibfield  {title} {\bibinfo {title} {Possible {Anomalies} at a {Semimetal}-{Semiconductor} {Transistion}},\ }\href {https://doi.org/10.1103/RevModPhys.40.755} {\bibfield  {journal} {\bibinfo  {journal} {Reviews of Modern Physics}\ }\textbf {\bibinfo {volume} {40}},\ \bibinfo {pages} {755} (\bibinfo {year} {1968})}\BibitemShut {NoStop}%
\bibitem [{\citenamefont {Halperin}\ and\ \citenamefont {{Rice, T.M.}}(1968)}]{halperin_excitonic_1968}%
  \BibitemOpen
  \bibfield  {author} {\bibinfo {author} {\bibfnamefont {B.~I.}\ \bibnamefont {Halperin}}\ and\ \bibinfo {author} {\bibnamefont {{Rice, T.M.}}},\ }\bibfield  {title} {\bibinfo {title} {The {Excitonic} {State} at the {Semiconductor}-{Semimetal} {Transition}},\ }\href {https://doi.org/10.1016/S0081-1947(08)60740-7} {\bibfield  {journal} {\bibinfo  {journal} {Journal of Physics C: Solid State Physics}\ }\textbf {\bibinfo {volume} {21}},\ \bibinfo {pages} {115} (\bibinfo {year} {1968})}\BibitemShut {NoStop}%
\bibitem [{\citenamefont {Volkov}\ and\ \citenamefont {Kopaev}(1973)}]{volkov_theory_1973}%
  \BibitemOpen
  \bibfield  {author} {\bibinfo {author} {\bibfnamefont {V.~A.}\ \bibnamefont {Volkov}}\ and\ \bibinfo {author} {\bibfnamefont {Y.~V.}\ \bibnamefont {Kopaev}},\ }\bibfield  {title} {\bibinfo {title} {Theory of phase transitions in semiconductors of the {A4}{B6} group},\ }\href@noop {} {\bibfield  {journal} {\bibinfo  {journal} {Zh. Eksp. Tear. Fiz}\ }\textbf {\bibinfo {volume} {64}},\ \bibinfo {pages} {2184} (\bibinfo {year} {1973})}\BibitemShut {NoStop}%
\bibitem [{\citenamefont {Jia}\ \emph {et~al.}(2022)\citenamefont {Jia}, \citenamefont {Wang}, \citenamefont {Chiu}, \citenamefont {Song}, \citenamefont {Yu}, \citenamefont {J{\"a}ck}, \citenamefont {Lei}, \citenamefont {Klemenz}, \citenamefont {Cevallos}, \citenamefont {Onyszczak}, \citenamefont {Fishchenko}, \citenamefont {Liu}, \citenamefont {Farahi}, \citenamefont {Xie}, \citenamefont {Xu}, \citenamefont {Watanabe}, \citenamefont {Taniguchi}, \citenamefont {Bernevig}, \citenamefont {Cava}, \citenamefont {Schoop}, \citenamefont {Yazdani},\ and\ \citenamefont {Wu}}]{jia_evidence_2022}%
  \BibitemOpen
  \bibfield  {author} {\bibinfo {author} {\bibfnamefont {Y.}~\bibnamefont {Jia}}, \bibinfo {author} {\bibfnamefont {P.}~\bibnamefont {Wang}}, \bibinfo {author} {\bibfnamefont {C.-L.}\ \bibnamefont {Chiu}}, \bibinfo {author} {\bibfnamefont {Z.}~\bibnamefont {Song}}, \bibinfo {author} {\bibfnamefont {G.}~\bibnamefont {Yu}}, \bibinfo {author} {\bibfnamefont {B.}~\bibnamefont {J{\"a}ck}}, \bibinfo {author} {\bibfnamefont {S.}~\bibnamefont {Lei}}, \bibinfo {author} {\bibfnamefont {S.}~\bibnamefont {Klemenz}}, \bibinfo {author} {\bibfnamefont {F.~A.}\ \bibnamefont {Cevallos}}, \bibinfo {author} {\bibfnamefont {M.}~\bibnamefont {Onyszczak}}, \bibinfo {author} {\bibfnamefont {N.}~\bibnamefont {Fishchenko}}, \bibinfo {author} {\bibfnamefont {X.}~\bibnamefont {Liu}}, \bibinfo {author} {\bibfnamefont {G.}~\bibnamefont {Farahi}}, \bibinfo {author} {\bibfnamefont {F.}~\bibnamefont {Xie}}, \bibinfo {author} {\bibfnamefont {Y.}~\bibnamefont {Xu}}, \bibinfo {author} {\bibfnamefont {K.}~\bibnamefont {Watanabe}}, \bibinfo
  {author} {\bibfnamefont {T.}~\bibnamefont {Taniguchi}}, \bibinfo {author} {\bibfnamefont {B.~A.}\ \bibnamefont {Bernevig}}, \bibinfo {author} {\bibfnamefont {R.~J.}\ \bibnamefont {Cava}}, \bibinfo {author} {\bibfnamefont {L.~M.}\ \bibnamefont {Schoop}}, \bibinfo {author} {\bibfnamefont {A.}~\bibnamefont {Yazdani}},\ and\ \bibinfo {author} {\bibfnamefont {S.}~\bibnamefont {Wu}},\ }\bibfield  {title} {\bibinfo {title} {Evidence for a monolayer excitonic insulator},\ }\href {https://doi.org/10.1038/s41567-021-01422-w} {\bibfield  {journal} {\bibinfo  {journal} {Nature Physics}\ }\textbf {\bibinfo {volume} {18}},\ \bibinfo {pages} {87} (\bibinfo {year} {2022})}\BibitemShut {NoStop}%
\bibitem [{\citenamefont {Sun}\ \emph {et~al.}(2022)\citenamefont {Sun}, \citenamefont {Zhao}, \citenamefont {Palomaki}, \citenamefont {Fei}, \citenamefont {Runburg}, \citenamefont {Malinowski}, \citenamefont {Huang}, \citenamefont {Cenker}, \citenamefont {Cui}, \citenamefont {Chu}, \citenamefont {Xu}, \citenamefont {Ataei}, \citenamefont {Varsano}, \citenamefont {Palummo}, \citenamefont {Molinari}, \citenamefont {Rontani},\ and\ \citenamefont {Cobden}}]{sun_evidence_2022}%
  \BibitemOpen
  \bibfield  {author} {\bibinfo {author} {\bibfnamefont {B.}~\bibnamefont {Sun}}, \bibinfo {author} {\bibfnamefont {W.}~\bibnamefont {Zhao}}, \bibinfo {author} {\bibfnamefont {T.}~\bibnamefont {Palomaki}}, \bibinfo {author} {\bibfnamefont {Z.}~\bibnamefont {Fei}}, \bibinfo {author} {\bibfnamefont {E.}~\bibnamefont {Runburg}}, \bibinfo {author} {\bibfnamefont {P.}~\bibnamefont {Malinowski}}, \bibinfo {author} {\bibfnamefont {X.}~\bibnamefont {Huang}}, \bibinfo {author} {\bibfnamefont {J.}~\bibnamefont {Cenker}}, \bibinfo {author} {\bibfnamefont {Y.-T.}\ \bibnamefont {Cui}}, \bibinfo {author} {\bibfnamefont {J.-H.}\ \bibnamefont {Chu}}, \bibinfo {author} {\bibfnamefont {X.}~\bibnamefont {Xu}}, \bibinfo {author} {\bibfnamefont {S.~S.}\ \bibnamefont {Ataei}}, \bibinfo {author} {\bibfnamefont {D.}~\bibnamefont {Varsano}}, \bibinfo {author} {\bibfnamefont {M.}~\bibnamefont {Palummo}}, \bibinfo {author} {\bibfnamefont {E.}~\bibnamefont {Molinari}}, \bibinfo {author} {\bibfnamefont {M.}~\bibnamefont {Rontani}},\ and\
  \bibinfo {author} {\bibfnamefont {D.~H.}\ \bibnamefont {Cobden}},\ }\bibfield  {title} {\bibinfo {title} {Evidence for equilibrium exciton condensation in monolayer {W}{Te}$_2$},\ }\href {https://doi.org/10.1038/s41567-021-01427-5} {\bibfield  {journal} {\bibinfo  {journal} {Nature Physics}\ }\textbf {\bibinfo {volume} {18}},\ \bibinfo {pages} {94} (\bibinfo {year} {2022})}\BibitemShut {NoStop}%
\bibitem [{\citenamefont {Que}\ \emph {et~al.}(2024)\citenamefont {Que}, \citenamefont {Chan}, \citenamefont {Jia}, \citenamefont {Das}, \citenamefont {Tong}, \citenamefont {Chang}, \citenamefont {Cui}, \citenamefont {Kumar}, \citenamefont {Singh}, \citenamefont {Mukherjee}, \citenamefont {Lin},\ and\ \citenamefont {Weber}}]{que_a-gate-tunable_2024}%
  \BibitemOpen
  \bibfield  {author} {\bibinfo {author} {\bibfnamefont {Y.}~\bibnamefont {Que}}, \bibinfo {author} {\bibfnamefont {Y.-H.}\ \bibnamefont {Chan}}, \bibinfo {author} {\bibfnamefont {J.}~\bibnamefont {Jia}}, \bibinfo {author} {\bibfnamefont {A.}~\bibnamefont {Das}}, \bibinfo {author} {\bibfnamefont {Z.}~\bibnamefont {Tong}}, \bibinfo {author} {\bibfnamefont {Y.-T.}\ \bibnamefont {Chang}}, \bibinfo {author} {\bibfnamefont {Z.}~\bibnamefont {Cui}}, \bibinfo {author} {\bibfnamefont {A.}~\bibnamefont {Kumar}}, \bibinfo {author} {\bibfnamefont {G.}~\bibnamefont {Singh}}, \bibinfo {author} {\bibfnamefont {S.}~\bibnamefont {Mukherjee}}, \bibinfo {author} {\bibfnamefont {H.}~\bibnamefont {Lin}},\ and\ \bibinfo {author} {\bibfnamefont {B.}~\bibnamefont {Weber}},\ }\bibfield  {title} {\bibinfo {title} {A gate-tunable ambipolar quantum phase transition in a topological excitonic insulator},\ }\href {https://doi.org/https://doi.org/10.1002/adma.202309356} {\bibfield  {journal} {\bibinfo  {journal} {Advanced Materials}\
  }\textbf {\bibinfo {volume} {36}},\ \bibinfo {pages} {2309356} (\bibinfo {year} {2024})}\BibitemShut {NoStop}%
\bibitem [{\citenamefont {Budich}\ \emph {et~al.}(2014)\citenamefont {Budich}, \citenamefont {Trauzettel},\ and\ \citenamefont {Michetti}}]{budich_time_2014}%
  \BibitemOpen
  \bibfield  {author} {\bibinfo {author} {\bibfnamefont {J.~C.}\ \bibnamefont {Budich}}, \bibinfo {author} {\bibfnamefont {B.}~\bibnamefont {Trauzettel}},\ and\ \bibinfo {author} {\bibfnamefont {P.}~\bibnamefont {Michetti}},\ }\bibfield  {title} {\bibinfo {title} {Time reversal symmetric topological exciton condensate in bilayer hgte quantum wells},\ }\href {https://doi.org/10.1103/PhysRevLett.112.146405} {\bibfield  {journal} {\bibinfo  {journal} {Phys. Rev. Lett.}\ }\textbf {\bibinfo {volume} {112}},\ \bibinfo {pages} {146405} (\bibinfo {year} {2014})}\BibitemShut {NoStop}%
\bibitem [{\citenamefont {Pikulin}\ and\ \citenamefont {Hyart}(2014)}]{pikulin_interplay_2014}%
  \BibitemOpen
  \bibfield  {author} {\bibinfo {author} {\bibfnamefont {D.~I.}\ \bibnamefont {Pikulin}}\ and\ \bibinfo {author} {\bibfnamefont {T.}~\bibnamefont {Hyart}},\ }\bibfield  {title} {\bibinfo {title} {Interplay of exciton condensation and the quantum spin hall effect in $\mathrm{InAs}/\mathrm{GaSb}$ bilayers},\ }\href {https://doi.org/10.1103/PhysRevLett.112.176403} {\bibfield  {journal} {\bibinfo  {journal} {Phys. Rev. Lett.}\ }\textbf {\bibinfo {volume} {112}},\ \bibinfo {pages} {176403} (\bibinfo {year} {2014})}\BibitemShut {NoStop}%
\bibitem [{\citenamefont {Xue}\ and\ \citenamefont {MacDonald}(2018)}]{xue_time-reversal_2018}%
  \BibitemOpen
  \bibfield  {author} {\bibinfo {author} {\bibfnamefont {F.}~\bibnamefont {Xue}}\ and\ \bibinfo {author} {\bibfnamefont {A.~H.}\ \bibnamefont {MacDonald}},\ }\bibfield  {title} {\bibinfo {title} {Time-reversal symmetry-breaking nematic insulators near quantum spin hall phase transitions},\ }\href {https://doi.org/10.1103/PhysRevLett.120.186802} {\bibfield  {journal} {\bibinfo  {journal} {Phys. Rev. Lett.}\ }\textbf {\bibinfo {volume} {120}},\ \bibinfo {pages} {186802} (\bibinfo {year} {2018})}\BibitemShut {NoStop}%
\bibitem [{\citenamefont {Blason}\ and\ \citenamefont {Fabrizio}(2020)}]{blason_exciton_2020}%
  \BibitemOpen
  \bibfield  {author} {\bibinfo {author} {\bibfnamefont {A.}~\bibnamefont {Blason}}\ and\ \bibinfo {author} {\bibfnamefont {M.}~\bibnamefont {Fabrizio}},\ }\bibfield  {title} {\bibinfo {title} {Exciton topology and condensation in a model quantum spin hall insulator},\ }\href {https://doi.org/10.1103/PhysRevB.102.035146} {\bibfield  {journal} {\bibinfo  {journal} {Phys. Rev. B}\ }\textbf {\bibinfo {volume} {102}},\ \bibinfo {pages} {035146} (\bibinfo {year} {2020})}\BibitemShut {NoStop}%
\bibitem [{\citenamefont {Varsano}\ \emph {et~al.}(2020)\citenamefont {Varsano}, \citenamefont {Palummo}, \citenamefont {Molinari},\ and\ \citenamefont {Rontani}}]{varsano_monolayer_2020}%
  \BibitemOpen
  \bibfield  {author} {\bibinfo {author} {\bibfnamefont {D.}~\bibnamefont {Varsano}}, \bibinfo {author} {\bibfnamefont {M.}~\bibnamefont {Palummo}}, \bibinfo {author} {\bibfnamefont {E.}~\bibnamefont {Molinari}},\ and\ \bibinfo {author} {\bibfnamefont {M.}~\bibnamefont {Rontani}},\ }\bibfield  {title} {\bibinfo {title} {A monolayer transition-metal dichalcogenide as a topological excitonic insulator},\ }\href {https://doi.org/10.1038/s41565-020-0650-4} {\bibfield  {journal} {\bibinfo  {journal} {Nature Nanotechnology}\ }\textbf {\bibinfo {volume} {15}},\ \bibinfo {pages} {367} (\bibinfo {year} {2020})}\BibitemShut {NoStop}%
\bibitem [{\citenamefont {Kwan}\ \emph {et~al.}(2021)\citenamefont {Kwan}, \citenamefont {Devakul}, \citenamefont {Sondhi},\ and\ \citenamefont {Parameswaran}}]{kwan_theory_2021}%
  \BibitemOpen
  \bibfield  {author} {\bibinfo {author} {\bibfnamefont {Y.~H.}\ \bibnamefont {Kwan}}, \bibinfo {author} {\bibfnamefont {T.}~\bibnamefont {Devakul}}, \bibinfo {author} {\bibfnamefont {S.~L.}\ \bibnamefont {Sondhi}},\ and\ \bibinfo {author} {\bibfnamefont {S.~A.}\ \bibnamefont {Parameswaran}},\ }\bibfield  {title} {\bibinfo {title} {Theory of competing excitonic orders in insulating {W}{Te}{$_2$} monolayers},\ }\href {https://doi.org/10.1103/PhysRevB.104.125133} {\bibfield  {journal} {\bibinfo  {journal} {Physical Review B}\ }\textbf {\bibinfo {volume} {104}},\ \bibinfo {pages} {125133} (\bibinfo {year} {2021})}\BibitemShut {NoStop}%
\bibitem [{\citenamefont {Amaricci}\ \emph {et~al.}(2023)\citenamefont {Amaricci}, \citenamefont {Mazza}, \citenamefont {Capone},\ and\ \citenamefont {Fabrizio}}]{amaricci_exciton_2023}%
  \BibitemOpen
  \bibfield  {author} {\bibinfo {author} {\bibfnamefont {A.}~\bibnamefont {Amaricci}}, \bibinfo {author} {\bibfnamefont {G.}~\bibnamefont {Mazza}}, \bibinfo {author} {\bibfnamefont {M.}~\bibnamefont {Capone}},\ and\ \bibinfo {author} {\bibfnamefont {M.}~\bibnamefont {Fabrizio}},\ }\bibfield  {title} {\bibinfo {title} {Exciton condensation in strongly correlated quantum spin hall insulators},\ }\href {https://doi.org/10.1103/PhysRevB.107.115117} {\bibfield  {journal} {\bibinfo  {journal} {Phys. Rev. B}\ }\textbf {\bibinfo {volume} {107}},\ \bibinfo {pages} {115117} (\bibinfo {year} {2023})}\BibitemShut {NoStop}%
\bibitem [{\citenamefont {Wang}\ \emph {et~al.}(2023)\citenamefont {Wang}, \citenamefont {Papaj},\ and\ \citenamefont {Moore}}]{wang_breakdown_2023}%
  \BibitemOpen
  \bibfield  {author} {\bibinfo {author} {\bibfnamefont {Y.-Q.}\ \bibnamefont {Wang}}, \bibinfo {author} {\bibfnamefont {M.}~\bibnamefont {Papaj}},\ and\ \bibinfo {author} {\bibfnamefont {J.~E.}\ \bibnamefont {Moore}},\ }\bibfield  {title} {\bibinfo {title} {Breakdown of helical edge state topologically protected conductance in time-reversal-breaking excitonic insulators},\ }\href {https://doi.org/10.1103/PhysRevB.108.205420} {\bibfield  {journal} {\bibinfo  {journal} {Phys. Rev. B}\ }\textbf {\bibinfo {volume} {108}},\ \bibinfo {pages} {205420} (\bibinfo {year} {2023})}\BibitemShut {NoStop}%
\bibitem [{\citenamefont {Papaj}(2024)}]{papaj_spectroscopic_2024}%
  \BibitemOpen
  \bibfield  {author} {\bibinfo {author} {\bibfnamefont {M.}~\bibnamefont {Papaj}},\ }\bibfield  {title} {\bibinfo {title} {Spectroscopic signatures of excitonic order on quantum spin {Hall} edge states},\ }\href {https://doi.org/10.1103/PhysRevB.110.165422} {\bibfield  {journal} {\bibinfo  {journal} {Phys. Rev. B}\ }\textbf {\bibinfo {volume} {110}},\ \bibinfo {pages} {165422} (\bibinfo {year} {2024})}\BibitemShut {NoStop}%
\bibitem [{\citenamefont {Amaricci}\ \emph {et~al.}(2015)\citenamefont {Amaricci}, \citenamefont {Budich}, \citenamefont {Capone}, \citenamefont {Trauzettel},\ and\ \citenamefont {Sangiovanni}}]{amaricci_first-order_2015}%
  \BibitemOpen
  \bibfield  {author} {\bibinfo {author} {\bibfnamefont {A.}~\bibnamefont {Amaricci}}, \bibinfo {author} {\bibfnamefont {J.~C.}\ \bibnamefont {Budich}}, \bibinfo {author} {\bibfnamefont {M.}~\bibnamefont {Capone}}, \bibinfo {author} {\bibfnamefont {B.}~\bibnamefont {Trauzettel}},\ and\ \bibinfo {author} {\bibfnamefont {G.}~\bibnamefont {Sangiovanni}},\ }\bibfield  {title} {\bibinfo {title} {First-order character and observable signatures of topological quantum phase transitions},\ }\href {https://doi.org/10.1103/PhysRevLett.114.185701} {\bibfield  {journal} {\bibinfo  {journal} {Phys. Rev. Lett.}\ }\textbf {\bibinfo {volume} {114}},\ \bibinfo {pages} {185701} (\bibinfo {year} {2015})}\BibitemShut {NoStop}%
\bibitem [{\citenamefont {Roy}\ \emph {et~al.}(2016)\citenamefont {Roy}, \citenamefont {Goswami},\ and\ \citenamefont {Sau}}]{roy_continuous_2016}%
  \BibitemOpen
  \bibfield  {author} {\bibinfo {author} {\bibfnamefont {B.}~\bibnamefont {Roy}}, \bibinfo {author} {\bibfnamefont {P.}~\bibnamefont {Goswami}},\ and\ \bibinfo {author} {\bibfnamefont {J.~D.}\ \bibnamefont {Sau}},\ }\bibfield  {title} {\bibinfo {title} {Continuous and discontinuous topological quantum phase transitions},\ }\href {https://doi.org/10.1103/PhysRevB.94.041101} {\bibfield  {journal} {\bibinfo  {journal} {Phys. Rev. B}\ }\textbf {\bibinfo {volume} {94}},\ \bibinfo {pages} {041101} (\bibinfo {year} {2016})}\BibitemShut {NoStop}%
\bibitem [{\citenamefont {Amaricci}\ \emph {et~al.}(2016)\citenamefont {Amaricci}, \citenamefont {Budich}, \citenamefont {Capone}, \citenamefont {Trauzettel},\ and\ \citenamefont {Sangiovanni}}]{amaricci_strong_2016}%
  \BibitemOpen
  \bibfield  {author} {\bibinfo {author} {\bibfnamefont {A.}~\bibnamefont {Amaricci}}, \bibinfo {author} {\bibfnamefont {J.~C.}\ \bibnamefont {Budich}}, \bibinfo {author} {\bibfnamefont {M.}~\bibnamefont {Capone}}, \bibinfo {author} {\bibfnamefont {B.}~\bibnamefont {Trauzettel}},\ and\ \bibinfo {author} {\bibfnamefont {G.}~\bibnamefont {Sangiovanni}},\ }\bibfield  {title} {\bibinfo {title} {Strong correlation effects on topological quantum phase transitions in three dimensions},\ }\href {https://doi.org/10.1103/PhysRevB.93.235112} {\bibfield  {journal} {\bibinfo  {journal} {Phys. Rev. B}\ }\textbf {\bibinfo {volume} {93}},\ \bibinfo {pages} {235112} (\bibinfo {year} {2016})}\BibitemShut {NoStop}%
\bibitem [{\citenamefont {K\"onig}\ \emph {et~al.}(2001)\citenamefont {K\"onig}, \citenamefont {Bonsager},\ and\ \citenamefont {MacDonald}}]{Konig_etal2001}%
  \BibitemOpen
  \bibfield  {author} {\bibinfo {author} {\bibfnamefont {J.}~\bibnamefont {K\"onig}}, \bibinfo {author} {\bibfnamefont {M.~C.}\ \bibnamefont {Bonsager}},\ and\ \bibinfo {author} {\bibfnamefont {A.~H.}\ \bibnamefont {MacDonald}},\ }\bibfield  {title} {\bibinfo {title} {Dissipationless spin transport in thin film ferromagnets},\ }\href@noop {} {\bibfield  {journal} {\bibinfo  {journal} {Phys. Rev. Lett.}\ }\textbf {\bibinfo {volume} {87}},\ \bibinfo {pages} {187202} (\bibinfo {year} {2001})}\BibitemShut {NoStop}%
\bibitem [{\citenamefont {Takei}\ and\ \citenamefont {Tserkovnyak}(2014)}]{Takei_Tserkovnyak2014}%
  \BibitemOpen
  \bibfield  {author} {\bibinfo {author} {\bibfnamefont {S.}~\bibnamefont {Takei}}\ and\ \bibinfo {author} {\bibfnamefont {Y.}~\bibnamefont {Tserkovnyak}},\ }\bibfield  {title} {\bibinfo {title} {Superfluid spin transport through easy-plane ferromagnetic insulators},\ }\href@noop {} {\bibfield  {journal} {\bibinfo  {journal} {Phys. Rev. Lett.}\ }\textbf {\bibinfo {volume} {112}},\ \bibinfo {pages} {227201} (\bibinfo {year} {2014})}\BibitemShut {NoStop}%
\bibitem [{\citenamefont {Takei}\ \emph {et~al.}(2016)\citenamefont {Takei}, \citenamefont {Yacoby}, \citenamefont {Halperin},\ and\ \citenamefont {Tserkovnyak}}]{Takei_etal2016}%
  \BibitemOpen
  \bibfield  {author} {\bibinfo {author} {\bibfnamefont {S.}~\bibnamefont {Takei}}, \bibinfo {author} {\bibfnamefont {A.}~\bibnamefont {Yacoby}}, \bibinfo {author} {\bibfnamefont {B.~I.}\ \bibnamefont {Halperin}},\ and\ \bibinfo {author} {\bibfnamefont {Y.}~\bibnamefont {Tserkovnyak}},\ }\bibfield  {title} {\bibinfo {title} {Spin superfluidity in the $\nu=0$ quantum hall state of graphene},\ }\href@noop {} {\bibfield  {journal} {\bibinfo  {journal} {Phys. Rev. Lett.}\ }\textbf {\bibinfo {volume} {116}},\ \bibinfo {pages} {216801} (\bibinfo {year} {2016})}\BibitemShut {NoStop}%
\end{thebibliography}%

\end{document}